\pdfoutput=1

\documentclass{article}
\usepackage{natbib,bm}
\bibpunct[; ]{(}{)}{,}{a}{}{;}

\usepackage{algorithm2e}
\usepackage{authblk}
\usepackage[utf8]{inputenc}
\usepackage{gensymb}
\usepackage{hyperref}
\usepackage{graphicx} %rotatebox
\usepackage{subcaption} %subfigure
\usepackage{amsmath, amssymb, bm}
\usepackage{units} %nicefrac
\usepackage{multirow}
\usepackage[a4paper]{geometry}
\usepackage[labelfont=bf]{caption}
\usepackage{float} %for begin{figure}[H]
\providecommand{\keywords}[1]
{\small	\textbf{Keywords: } #1 }

\usepackage[flushmargin]{footmisc}

\def\bfG{\mathbf G}

\def\bfI{\mathbf I}

\def\bfQ{\mathbf Q}

\def\bfS{\mathbf S}

\def\bfV{\mathbf V}
\def\bfW{\mathbf W}
\def\bfX{\mathbf X}
\def\bfY{\mathbf Y}

\def\bfb{\mathbf b}

\def\bfe{\mathbf e}

\def\bfy{\mathbf y}

\def\bfbeta   {\bm \beta}

\def\bfepsilon{\bm \epsilon}

\def\bfSigma  {\mathbf \Sigma}

\newcommand{\bfzero}{{\mathbf 0}}

\def\boldfacefake#1{\kern-4pt
    \hbox{ \mathsurround=0pt
    \hbox to 0.4pt{$#1$\hss}\hbox to 0.4pt{$#1$\hss}\hbox {$#1$}}}

\newcommand{\be}{\begin{eqnarray}}
\newcommand{\ee}{\end{eqnarray}}
\newcommand{\ba}{\begin{eqnarray*}}
\newcommand{\ea}{\end{eqnarray*}}
\newcommand{\bc}{\begin{center}}
\newcommand{\ec}{\end{center}}
\newcommand{\btab}[1]{\begin{tabular}{#1}}
\newcommand{\etab}{\end{tabular}}

\newcommand{\reals}{\mbox{\rm I\kern-.20em R}}
\newcommand{\sreals}{\mbox{\small \rm I\kern-.20em R}}

\title{Sources of residual autocorrelation in multiband task fMRI and strategies for effective mitigation}

\author[1]{Fatma Parlak} 
\author[1]{Damon D. Pham}
\author[1]{Daniel A. Spencer}
\author[2]{Robert C. Welsh}
\author[1]{Amanda F. Mejia \thanks{Corresponding author: Amanda F. Mejia, mandy.mejia@gmail.com}} 
\date{}

\affil[1]{Department of Statistics, Indiana University, Bloomington, IN, USA}

\affil[2]{Department of Psychiatry and Biobehavioral Sciences, Los Angeles, CA, USA}

\begin{document}

\maketitle

\begin{abstract}

Analysis of task fMRI studies is typically based on using ordinary least squares within a voxel- or vertex-wise linear regression framework known as the general linear model. This use produces estimates and standard errors of the regression coefficients representing amplitudes of task-induced activations. To produce valid statistical inferences, several key statistical assumptions must be met, including that of independent residuals. Since task fMRI residuals often exhibit temporal autocorrelation, it is common practice to perform ``prewhitening'' to mitigate that dependence. Prewhitening involves estimating the residual correlation structure and then applying a filter to induce residual temporal independence. While theoretically straightforward, a major challenge in prewhitening for fMRI data is accurately estimating the residual autocorrelation at each voxel or vertex of the brain. Assuming a global model for autocorrelation, which is the default in several standard fMRI software tools, may under- or over-whiten in certain areas and produce differential false positive control across the brain. The increasing popularity of multiband acquisitions with faster temporal resolution increases the challenge of effective prewhitening because more complex models are required to accurately capture the strength and structure of autocorrelation. These issues are becoming more critical now because of a trend towards subject-level analysis and inference. In group-average or group-difference analyses, the within-subject residual correlation structure is accounted for implicitly, so inadequate prewhitening is of little real consequence.  For individual subject inference, however, accurate prewhitening is crucial to avoid inflated or spatially variable false positive rates. In this paper, we first thoroughly examine the patterns, sources and strength of residual autocorrelation in multiband task fMRI data.  We find that residual autocorrelation exhibits marked spatial variance across the cortex and is influenced by many factors including the task being performed, the specific acquisition protocol, mis-modeling of the hemodynamic response function, unmodeled noise due to subject head motion, and systematic individual differences. Second, we evaluate the ability of different autoregressive (AR) model-based prewhitening strategies to effectively mitigate autocorrelation and control false positives. We consider two main factors: the choice of AR model order and the level of spatial regularization of AR model coefficients, ranging from local smoothing to global averaging. We also consider determining the AR model order optimally at every vertex, but we do not observe an additional benefit of this over the use of higher-order AR models (e.g. (AR(6)). We find that local regularization is much more effective than global averaging at mitigating autocorrelation. While increasing the AR model order is also helpful, it has a lesser effect than allowing AR coefficients to vary spatially. We find that prewhitening with an AR(6) model with local regularization is effective at reducing or even eliminating autocorrelation and controlling false positives. To overcome the computational challenge associated with spatially variable prewhitening, we developed a computationally efficient R implementation using parallelization and fast C++ backend code. This implementation is included in the open source R package \texttt{BayesfMRI}.
\end{abstract}

%Intro:
%The Olszowy paper and others have shown us that existing prewhitening methods are not sufficient. The Corbin paper also showed this.  There was also a paper that looked at varying AR model order across the brain, and that helped a lot.

\section{Introduction}

The general linear model (GLM) has long been a popular framework for the analysis of task functional magnetic resonance imaging (fMRI) data. In the GLM, a linear regression model is used to relate the observed blood oxygenation level dependent (BOLD) signal to the expected BOLD response due to each task or stimulus in the experiment, along with nuisance regressors, yielding an estimate of the activation across the brain due to each task \citep{friston1994statistical}. Hypothesis testing with multiplicity correction is used to determine areas of the brain that are significantly activated in response to each task or contrast. One well-known issue with the GLM approach is that BOLD data generally violates the ordinary least squares (OLS) assumption of residual independence \citep{Lindquist_2008, monti2011statistical}. When this happens, standard errors associated with the model coefficients are biased, invalidating inference and generally giving rise to inflated false positive rates for areas of activation. 

Violations of residual independence are most consequential for subject-level inference (the ``first-level GLM''), since OLS-based group inference (the ``second-level GLM'') has been found to be relatively robust to dependent errors in the first level \citep{mumford2009simple}. While group-level analysis has historically been the norm in fMRI studies, more recently subject-level analysis is gaining in relevance. This in part due to the rise of ``highly sampled'' datasets collecting lots of data on individual subjects \citep{laumann2015functional, choe2015reproducibility, braga2017parallel, gordon2017precision}, as well as growing interest in using fMRI data for biomarker discovery, clinical translation, and other contexts where robust and reliable subject-level measures are required. Unfortunately, fMRI data presents several challenges for proper statistical analysis \citep{monti2011statistical}, and subject-level task fMRI measures have been shown to be unreliable \citep{elliott2020test}. One important factor for reliable subject-level task fMRI analysis is dealing appropriately with temporal dependence to avoid inflated rates of false positives.

Generalized least squares (GLS) is a regression framework that accounts for dependent and/or heteroskedastic errors. Briefly, in a regression model $\bfy = \bfX\bfbeta + \bfepsilon$, $\bfepsilon\sim N(\bfzero, \bfV)$, assume that the residual covariance matrix $\mathbf{V}$ is known. Then the GLS solution is $\hat\bfbeta^{GLS} = (\bfX'\bfW\bfX)^{-1}\bfX'\bfW\bfy$, where $\bfW=\bfV^{-1}$, and $Var(\hat\bfbeta^{GLS})=(\bfX'\bfW\bfX)^{-1}$. This is mathematically equivalent to pre-multiplying both sides of the regression equation by $\bfV^{-1/2}$, which induces independent and homoskedastic residuals and gives rise to an OLS solution of the same form as $\hat\bfbeta^{GLS}$. In the context of task fMRI analysis, such ``prewhitening'' is a common remedy to eliminate temporal dependence as it produces the best linear unbiased estimate (BLUE) of model coefficients \citep{Bullmore_1996}. A key challenge in GLS analysis is determining the form of $\bfV$, which is not actually known in practice.  In a conventional statistical analysis, an iterative approach to estimating $\bfV$ is commonly used through iteratively reweighted least squares (IRLS). However, in fMRI analysis this is typically considered computationally prohibitive and prone to overfitting, so $\bfV$ is often estimated in a single pass based on the OLS residuals \citep{woolrich2001temporal}. To constrain the estimation of $\bfV$, a parametric form is typically assumed based on the temporal structure of the data, such as an autoregressive (AR) or AR moving average (ARMA) model. It is also common practice to regularize the AR or ARMA model parameters by smoothing or averaging across the brain or within tissue boundaries.

Prewhitening methods are implemented in the major software packages AFNI \citep{cox1996afni}, FSL \citep{jenkinson2012fsl} and SPM \citep{penny2011statistical}. Yet, many of these standard prewhitening techniques have received criticism for failing to effectively remove residual autocorrelation \citep{worsley2002general, Eklund_2012}.  These criticisms have pointed to two main sources of mismodelled residual autocorrelation: (1) use of overly parsimonious autocorrelation models that fail to fully capture the autocorrelation in the data---an issue of ever-increasing relevance with the rise of faster multi-band acquisitions---and/or (2) assuming the same degree of residual autocorrelation across the brain. \cite{olszowy2019accurate} performed a systematic comparison of prewhitening techniques implemented in SPM, AFNI and FSL using several task and rest datasets of varying repetition time (TR) between 0.645s and 3s. They found that, while some techniques clearly performed better than others, all failed to control false positives at the nominal level, especially for low TR data. The best performance was seen using AFNI, which assumes a first-order autoregressive moving average (ARMA) model with unsmoothed, spatially varying coefficients, and the FAST option in SPM, which employs a flexible but global model using a dictionary of covariance components \citep{corbin2018accurate}. Interestingly, these two methods represent opposite approaches: AFNI allows for spatially varying autocorrelation but uses a relatively restrictive ARMA(1,1) model, while SPM FAST uses a quite flexible temporal correlation model but imposes a restrictive global assumption.  Neither AFNI, FSL or SPM currently offers a prewhitening technique that provides for a flexible \textit{and} spatially varying autocorrelation model. Therefore, the ability to fully account for residual autocorrelation remains a limitation of many first-level task fMRI analyses.

Several recent studies have considered the ability of higher-order autoregressive models to adequately capture residual autocorrelation in fast TR fMRI data \citep{bollmann2018serial, chen2019analysis, luo2020improved}.  \cite{bollmann2018serial} found that optimal AR model order and AR coefficient magnitude varied markedly across the brain in fast TR task fMRI data, and that physiological noise modeling reduced but by no means eliminated the spatial variability in residual autocorrelation or the need for a high AR model order. \cite{luo2020improved} used resting-state fMRI data of varying sub-second TRs to examine false positives rates with assumed task paradigms. They found that the optimal AR model order varied spatially and depended on TR, with faster TR requiring a higher AR model order. They found that a too-low or too-high AR model order resulted in inflated false positive rates, and that a global model order (even with spatially varying coefficients) performed worse than when model order was allowed to vary spatially. Their approach of allowing the AR model order to vary spatially also outperformed both SPM FAST and the ARMA(1,1) model used by AFNI, the two methods found to have the best performance by \cite{olszowy2019accurate}.

These recent studies also highlight several challenges associated with the use of volumetric fMRI in prewhitening. Both \cite{bollmann2018serial} and \cite{luo2020improved} observed sharp differences in the strength of residual autocorrelation across tissue classes, with cerebral spinal fluid (CSF) exhibiting much stronger autocorrelation than gray matter, and white matter exhibiting relatively low autocorrelation. Because of this, they point out that the standard practice of spatial smoothing (of the data, of the AR model order, or of the AR coefficient estimates) may be problematic at tissue class boundaries: gray matter bordering CSF may have higher autocorrelation due to mixing with CSF signals, while gray matter bordering white matter may have decreased autocorrelation due to mixing with white matter signals.  Indeed, \cite{luo2020improved} found smoothing of the sample autocorrelations at 6mm FWHM to result in inflated false positive rates.  Yet some regularization of autocorrelation model parameters is believed to be necessary to avoid very noisy estimates \citep{worsley2002general, bollmann2018serial, chen2019analysis}, and data smoothing is nearly universal practice in the massive univariate framework, given its ability to enhance signal-to-noise ratio (SNR) and increase power to detect activations.  This presents a dilemma: smoothing across tissue classes can be detrimental for autocorrelation modeling, but regularization of autocorrelation coefficients is needed to avoid overly noisy estimates.

The use of cortical surface fMRI (cs-fMRI) could mitigate this dilemma in two ways. First, geodesic smoothing along the surface can increase SNR without blurring across tissue classes or neighboring sulcal folds. Second, by eliminating white matter and CSF, the spatial variability in residual autocorrelation is simplified, since the most dramatic spatial differences have been observed between tissue classes.  An additional potential benefit of the use of cs-fMRI is the utility of spatial Bayesian models, which cs-fMRI is uniquely suited for \citep{mejia2020bayesian}, to spatially regularize autocorrelation coefficients in a statistically principled way. Therefore, in this work we adopt cortical surface format fMRI.

In this work, we advance prewhitening methods for modern fMRI acquisitions in three ways. First, we thoroughly examine the spatial variability and influence of various factors on residual autocorrelation, including the task protocol, the acquisition technique, and systematic individual differences. We also examine the influence of potential model mis-specification for the GLM. Un-modelled neural activity is temporally correlated and may be absorbed into the model residuals, thus increasing residual autocorrelation \citep{lindquist2009modeling, bollmann2018serial}.  %In the basic form of the classical GLM, a canonical hemodynamic response function (HRF) is convolved with a . Because 
Task-related neural activity is assumed to be captured through task regressors, which are typically constructed by convolving a hemodynamic response function (HRF) with a stick function  representing the task paradigm. However, the shape and duration of the HRF are known to vary across the brain \citep{Lindquist_2007} as well as within and across individuals \citep{aguirre1998variability}. Therefore, assuming a fixed ``canonical'' HRF may fail to adequately capture task-induced activity \citep{lindquist2009modeling}. More flexible models can capture differences in HRF height, width, and time to peak, including the model with temporal and dispersion derivatives \citep{friston1998nonlinear}, the finite impulse response model \citep{glover1999deconvolution}, and the inverse logit model \citep{Lindquist_2007}. Here, we consider the effect of including the temporal and/or dispersion derivatives of the HRF on autocorrelation. Another potential source of model mis-specification is failure to account for noise resulting from head motion, scanner drift, and other sources. If such noise is not modeled, it will be reflected in the model residuals. Because such sources of noise tend to be temporally correlated, this will tend have the effect of increasing residual autocorrelation.  Here, we consider the effect of including more or fewer head motion-based regressors on residual autocorrelation.

Second, we evaluate the effectiveness of different autoregressive (AR) model-based prewhitening strategies at reducing autocorrelation and controlling false positive rates. We consider AR model order varying from $1$ to $6$, as well as determining the AR model order optimally at each vertex. We also consider local regularization of AR model coefficients versus global averaging. We find that local surface-based regularization of AR model coefficients is much more effective than a global prewhitening strategy at eliminating autocorrelation across the cortex. 

Third, we overcome the major computational challenges associated with spatially-varying prewhitening. We have developed a computationally efficient implementation of the AR-based prewhitening techniques considered here. Using parallelization and backend code written in C++, we are able to perform spatially varying prewhitening very efficiently for surface-based analysis and ``grayordinates'' analysis more generally. This implementation is available in the open-source \texttt{BayesfMRI} R package \citep{mejia2022BayesfMRI}.

The remainder of this paper is organized as follows.  In Section \ref{sec:methods} we describe the data, the GLM approach, and the methods for autocorrelation estimation, and prewhitening. We also describe a mixed effect modeling framework we use to assess the influence of several key factors on the strength and spatial variability of residual autocorrelation, including acquisition method, task protocol, modeling choices, and individual variability. In Section \ref{sec:results} we present results based on an analysis of several task and resting state fMRI studies from the Human Connectome Project, utilizing 40 subjects with test-retest data. In Section \ref{sec:discussion}, we conclude with a discussion of these results and what they suggest for future research in prewhitening.

%NEW PAPERS TO READ (BASED ON THOSE THAT CITED OLSZOWY PAPER):

% @article{demetriou2018comprehensive,
%   title={A comprehensive evaluation of increasing temporal resolution with multiband-accelerated protocols and effects on statistical outcome measures in fMRI},
%   author={Demetriou, Lysia and Kowalczyk, Oliwia S and Tyson, Gabriella and Bello, Thomas and Newbould, Rexford D and Wall, Matthew B},
%   journal={Neuroimage},
%   volume={176},
%   pages={404--416},
%   year={2018},
%   publisher={Elsevier}
% }

\section{Materials and methods}
\label{sec:methods}

\subsection{Data Collection and Processing}
The data used in this paper are from the Human Connectome Project (HCP) 1200-subject release \\(http://humanconnectome.org). The HCP includes task and resting-state fMRI data collected on a customized Siemens 3T Skyra scanner with a multiband factor of 8 to provide high spatial (2mm isotropic voxels) and temporal (TR = 0.72s) resolution \citep{van2013wu}. The fMRI data were processed according to the HCP minimal preprocessing pipelines including projection to the cortical surface, as described in \cite{glasser2013minimal}. The resulting surface mesh for each hemisphere consists of approximately 32,000 vertices. For all fMRI scans, we perform surface-based spatial smoothing using a 2-dimensional Gaussian kernel with $6$mm full-width-at-half-maximum (FWHM). To reduce the computational burden of estimating the mixed-effects models described below, prior to smoothing, we resample to approximately 6,000 vertices per hemisphere. Note that this level of resampling results in a much milder degree of interpolation than smoothing at 6mm FWHM, and therefore results in a negligible loss of information when performed in combination with smoothing (see, e.g., \cite{mejia2020bayesian} Appendix Figure C4). For both resampling and smoothing, we employ the Connectome Workbench \citep{marcus2011informatics} via the \texttt{ciftiTools} R package \citep{pham2022ciftitools}. 

Each subject underwent several task and resting-state fMRI protocols across two sessions. Each task and rest session was performed twice, using opposite phase encoding directions (LR and RL). For a subset of 45 participants, the entire imaging protocol was repeated.  We analyze data from the 40 participants having a complete set of test and retest data for the protocols we analyze in this study. We analyze four task experiments, namely the emotion, gambling, motor, and relational tasks (Table \ref{tab:tasks}).  In the emotion processing task, developed by \cite{hariri_amygdala_2002}, participants are shown sets of faces or geometric shapes, and are asked to determine which of two faces/shapes match a reference face/shape. Each face has an angry or fearful expression.  A 3s cue  ("shape" or "face") precedes a block of 6 trials, lasting 18s in total. Each run includes three blocks of each condition (shape or face).

In the gambling task, adopted from \cite{Delgado2000TrackingTH}, participants play a game in which they are asked to guess the value of a mystery card to win or lose money.  They indicate their guess for the value, which can range from 1 to 9, as being more or less than 5. Their response is evaluated by a program which predetermines whether the trial is a win, loss, or neutral event. In each run, there are 2 mostly win blocks (6 win trials and 2 non-win trials), 2 mostly loss blocks (6 loss trials and 2 non-loss trials), alternating with 4 15s fixation blocks. While this protocol can be considered a block design since the task protocol is comprised of short events rather than continuous blocks of stimulus, we analyze it as an event-related design.

In the motor task, developed by \cite{buckner_organization_2011}, participants are given a 3s visual cue which instructs them to perform one of five motor tasks: tap left or right fingers, wiggle left or right toes, or move tongue. Each task block lasts 12 seconds, and each run includes two blocks of each task as well as three 15s fixation blocks.   

In the relational task, developed by \cite{smith_localizing_2007}, subjects undergo two conditions: relational processing and control matching. In the relational condition, one pair of objects is shown at the top of the screen and another pair is shown at the bottom of the screen. In this condition, participants are asked to determine the dimension (shape or texture) across which the pair displayed at the top differs. Next, they determine whether the bottom pair differs along the same dimension. During the matching condition, two objects are displayed at the top of the screen and one is shown at the bottom, and the word "shape" or "texture" appears in the middle of the screen. In this condition, participants are asked to determine whether the bottom object matches either of the top objects, based on the dimension displayed in the middle. Each condition is administered as blocks of trials of the same condition, with each block lasting 18s total, with three blocks of each condition (relational, matching and fixation block) in each run. 

To quantify false positive rates, we also analyze resting-state fMRI data acquired for the same subjects and sessions, which we analyze under a false task protocol.  To emulate the duration of the task fMRI runs, we truncate the resting-state runs to have $284$ volumes, the same number of volumes as the motor task, after dropping the first $15$ rest volumes. The boxcar design consists of a single event with three ``boxcars": three periods of stimulus lasting ten seconds each, with ten seconds in between each consecutive stimulus. The first stimulus begins at the 20th second, or approximately the 28th volume. We use only three boxcars instead of extending the boxcars to the duration of the scan, in order to more closely resemble the number of stimuli in the HCP task scans. We used the same GLM model as with the task analysis, except we did not include any HRF derivatives since there is no true task-evoked signal to potentially mis-model. 

% Table 1
\begin{table}[H]
     \caption{Type and duration of each task protocol analyzed. Adopted from \cite{Barch_2013}.}
    \centering
    \begin{tabular}{l|c|c|c}
    \hline
    Task & Block or Event & Frames per run & Run duration (mm:ss) \\
    \hline
        Emotion & Block & 176 & 2:16 \\
        Gambling & Event & 253 & 3:12 \\
        Motor & Block & 284 & 3:34 \\
        Relational & Block & 232 & 2:56 \\
    \hline
    \end{tabular}
    \label{tab:tasks}
\end{table}

\subsection{Statistical Analysis}

Our analysis consists of three primary steps. First, for each subject, session, task, acquisition protocol, HRF modeling strategy, and motion regression strategy, we fit a vertex-wise general linear model (GLM) to estimate the amplitude of task-evoked activation assuming residual independence. Based on the fitted residuals, we estimate the degree of autocorrelation at every location in the brain. Second, we fit a series of mixed effects models to identify the effects of acquisition and modeling factors on residual autocorrelation across the brain, as well as systematic individual variability.  Finally, we prewhiten the data using a range of strategies, varying the parametric model order and spatial regularization level. We evaluate the ability of each prewhitening strategy to effectively mitigate autocorrelation and control false positives.

\subsubsection{GLM Estimation}
\label{sec:GLM}

We first fit a series of GLMs to each task fMRI dataset assuming residual independence in order to quantify residual autocorrelation and examine its patterns and sources. Let $\bfy_v$ be the BOLD response at vertex $v$, and let $\bfX$ be a design matrix containing an intercept, task-related regressors, and nuisance regressors.  For each vertex $v$, the GLM proposed by \cite{friston1994statistical} can be written as:
\begin{align}\label{eqn:GLM}
    \bfy_v= \bfX\bfbeta_v +\bfepsilon_v,\quad
    \bfepsilon_v \sim MVN(\bfzero, \bfSigma_v).
\end{align}

$\bfSigma_v$ $(T\times T)$ encodes the residual autocorrelation and variance, which may differ across the brain. If $\bfSigma_v \ne \sigma^2_v\bfI$, the OLS assumption of residual independence is violated, and a generalized least squares (GLS) approach is appropriate in place of OLS to improve estimation efficiency and to avoid invalid statistical inference. In some cases, such as in spatial Bayesian variants of the GLM where a single Bayesian linear model is fit to all vertices, spatially homogeneous variance may also be assumed, i.e. $\sigma^2_v\equiv \sigma^2$.  GLS can be used to satisfy this assumption by inducing unit variance across the brain. In GLS, OLS is first used to obtain an initial set of fitted residuals $\hat{\bfepsilon}_v$, which are utilized to estimate $\bfSigma_v$ \citep{kariya2004generalized}. The GLS coefficient estimates are given by $\hat\bfbeta_v^{GLS}=(\bfX'\bfSigma_v^{-1}\bfX)^{-1}\bfX'\bfSigma_v^{-1}\bfy$, and their covariance is $Var(\hat\bfbeta_v^{GLS})=(\bfX'\bfSigma_v^{-1}\bfX)^{-1}$. Equivalently, prewhitening involves pre-multiplying both sides of the regression equation (\ref{eqn:GLM}) by $\bfW_v = \bfSigma_v^{-1/2}$ to induce residual independence. Traditionally this process is repeated until convergence, but due to computational considerations a single iteration is often assumed to be sufficient for task fMRI analysis \citep{woolrich2001temporal}. 

To avoid overly noisy estimates of $\bfSigma_v$, restrictive parametric models (e.g., low-order autoregressive models) and/or aggressive regularization (e.g., averaging across all gray matter) are often used to estimate $\bfSigma_v$.  Recent work has suggested that these approaches generally fail to fully account for autocorrelation or control false positives as a nominal rate \citep{luo2020improved, olszowy2019accurate, chen2019analysis, corbin2018accurate, bollmann2018serial}. The challenge is how to produce a sufficiently efficient estimate of $\bfSigma_v$ while accurately representing the differential levels of autocorrelation across the brain. In Section \ref{sec:prewhitening}, we consider various strategies for estimation of $\bfW_v = \bfSigma_v^{-1/2}$. However, our first step is to examine the sources and patterns of residual autocorrelation, and therefore we do not impose any parametric model or regularization in estimating $\bfSigma_v$. Instead, we use empirical autocorrelation function (ACF) at each vertex.  Consider the timeseries of fitted OLS residuals $\bfe_v = \bfy_v - \bfX(\bfX'\bfX)^{-1}\bfX'\bfy$ at vertex $v$ for a particular fMRI dataset. The ACF of $\bfe_v$ at lag $u$ is defined as 
$$
\rho_{v,u} = \frac{Cov(e_{v,t}, e_{v,t+u})}{\sqrt{Var(e_{v,t})Var(e_{v,t+u})}},
$$
for $u=0,\dots,T-1$, with lag-0 ACF $\rho_{v,0}=1$ \citep{venables2013modern}. We summarize the ACF $\rho_{v,u}$ into a single metric of autocorrelation, the autocorrelation index (ACI) \citep{afyouni2019effective}, which is given by
$$
\tau(v) = \sum_{u=0}^{T-1} \rho_{u}^2(v).
$$
%We also aim to examine the spatial variability of the residual variance $\sigma^2_v$ and the efficacy of prewhitening to homogenize the variance across the brain. To estimate $\sigma^2_v$, we assume $\bfSigma_v$ follows an autoregressive process. We fit a high-order autoregressive model with no spatial regularization to avoid inducing bias. Specifically, $\bfe_v$ is assumed to follow a zero-mean AR($p$) process:
% \begin{align}\label{eqn:AR}
%     e_v(t) = \sum_{r=1}^{p}\phi_{v,r}\,e_v(t-r) \, + w_v(t),
% \end{align}
% where $\phi_{v,r}$ is the $r^{th}$ AR coefficient and $w_v(t)\stackrel{iid}{\sim} N(0,\sigma^2_v)$ is a white noise process. Using the Yule-Walker equations, we obtain estimates of the AR coefficients and of $\sigma^2_v$ \citep{Brockwell_Davis}. 

We consider the effect of two potential sources of model misspecification on temporal autocorrelation: unmodeled neuronal activity via the task regressors in $\bfX$, and unmodeled head motion-induced noise via the nuisance regressors in $\bfX$.  Since both neuronal activity and motion-induced noise exhibit temporal dependence, failing to adequately account for either may contribute to residual autocorrelation.  The task regressors in $\bfX$ are constructed by convolving a stimulus function representing the timing of the tasks or stimuli with a canonical HRF, which is typically modeled as a gamma function or a difference of two gamma functions \citep{worsley2002general}. However, HRF onset and duration is known to vary across the brain and across individuals \citep{aguirre1998variability}, so using a fixed HRF may fail to accurately capture the task-evoked BOLD signal \citep{Ji_Loh_2008, lindquist2009modeling}. Therefore, we consider three models for the HRF: one assuming a fixed canonical HRF; one including the temporal derivative of the HRF to allow for differences in HRF onset timing; and one additionally including its dispersion derivative to allow for differences in HRF duration \citep{friston1998event, lindquist2009modeling}. 
Regarding nuisance regressors, the inclusion of measures of head motion is a common practice to account for head motion-induced noise in the data. We therefore consider two sets of motion regressors: the 6 rigid body realignment parameters and their one-back differences (RP12) or those terms plus their squares (RP24). In all models, we include discrete cosine transform (DCT) bases to achieve high-pass filtering at 0.01 Hz, which is important to satisfy the stationarity assumption of AR-based prewhitening.

In sum, we estimate a vertex-wise GLM via OLS for each subject $i=1,\dots,40$, session $j=1,2$, task $k=1,2,3,4$, and phase encoding direction $\in\{\text{LR,RL}\}$.  Each GLM is fit using the canonical HRF only, with its temporal derivative, and with its temporal and dispersion derivatives. Each GLM is also fit with 12 or 24 motion realignment parameters. In total, we fit 3,840 GLMs ($40\times2\times4\times2\times3\times2$) before prewhitening. All models are fit using the \texttt{BayesfMRI} R package \cite{mejia2022BayesfMRI}.  In the next section, we describe the mixed effects modeling framework we use to disentangle the influence of each factor (e.g. subject effects versus acquisition effects versus modeling effects) on residual autocorrelation.

%12 mp = r= 0 .  For each scenario, we quantify at each vertex the autocorrelation index $\tau_v$ and residual variance $\sigma^2_v$.  In the next section, we describe the mixed effects modeling framework we use to assess the spatial variability and influence of various factors on both autocorrelation and residual variance.

\subsubsection{Examining Sources of Residual Autocorrelation through Mixed Effects Modeling}
\label{sec:ACI}

Let $\tau_{ijk}^{hr\ell}(v)$ be the ACI at vertex $v$ for subject $i$, session $j$, task $k$, phase encoding direction $\ell$, HRF modeling strategy $h$ and motion regression strategy $r$.  To determine the influences of population variability, spatial variability and other factors on ACI, we fit a mixed effect model at each vertex.  We include fixed effects for each task, for the interaction between task and HRF modeling strategy, and for the motion regression strategy. For each of these fixed effects, we also include a random effect to represent population heterogeneity. Finally, we include a fixed effect for phase encoding direction. The mixed effect model at vertex $v$ is given by:

\begin{equation}\label{eqn:lmer_model}
    \tau_{ijk}^{hr\ell}(v) = 
    \overbrace{
    \left\{\alpha_k(v) + a_{k,i}(v)\right\}}^\text{baseline effects}\ +\  
    \overbrace{\left\{\beta_k(v) + b_{k,i}(v)\right\}h}^\text{HRF modeling effects}\ +\  
    \overbrace{\left\{\gamma(v) + g_i(v) \right\}r}^{\substack{\text{motion} \\ \text{regression effects}}}\ +
    \overbrace{\theta(v) \ell} ^{\substack{\text{acquisition}\\\text{effects}}}+\ 
    \epsilon_{ijk}^{hr\ell}(v)
\end{equation}
\begin{equation*}
    \epsilon_{ijk}^{hr\ell}(v) \stackrel{iid}{\sim} N(0,\sigma^2_v),
    \bfb_{i}(v) = \left\{a_{1,i}(v),\dots,a_{4,i}(v),b_{1,i}(v),\dots,b_{4,i}(v),g_i(v)\right\}' \stackrel{iid}{\sim} N(\bfzero,\bfG(v)).
\end{equation*}

The covariates $h$, $r$ and $\ell$ are constructed as dummy variables equalling zero for the ``baseline'' conditions (canonical HRF only, RP12 motion regression, and LR phase encoding direction acquisition) and equalling one for the alternative conditions (canonical HRF plus derivative(s), RP24 motion regression, and RL phase encoding direction acquisition). The model in equation (\ref{eqn:lmer_model}) is estimated separately for $h=1$ representing the inclusion of HRF temporal derivatives or the HRF temporal and dispersion derivatives.

We perform model fitting using the \texttt{lmer} function from the \texttt{lme4} R package (version 1.1.-30) \citep{lme4} to estimate each fixed effect (the $\alpha_k(v)$, $\beta_k(v)$, $\gamma(v)$ and $\theta(v)$), the error variance, and the covariance of all the random effects (the $a_{k,i}(v)$, $b_{k,i}(v)$ and $g_i(v)$). The fixed effect for task, $\alpha_k(v)$, represents the baseline autocorrelation for task protocol $k$ for the model including the canonical HRF only ($h=0$). The corresponding random subject effect, $a_{k,i}(v)$, represents the difference in autocorrelation for subject $i$, versus the average over subjects for task $k$. The fixed effect for HRF modeling, $\beta_k(v)$, represents the change in autocorrelation when the temporal derivative of the HRF is included ($h=1$) for task $k$, while $b_{k,i}(v)$ represents the random variation in that change over subjects. We generally expect negative values for $\beta_k(v)$, representing a reduction in residual autocorrelation when the HRF derivative is included, since discrepancies between the true HRF and the canonical HRF tend to exhibit temporal dependence. The fixed effect for motion regression strategy, $\gamma(v)$, represents the change in autocorrelation associated with the use of RP24 ($r=1$) versus RP12 ($r=0$) motion regression. The corresponding random effect, $g_i(v)$, represents random variation in that effect over subjects. $\theta(v)$ represents the difference in autocorrelation when using phase encoding direction $RL$ ($\ell=1$), compared with phase encoding direction $LR$ ($\ell=0$).  

%here
Since the model includes multiple sessions from each subject, the random effects $a_{k,i}(v)$, $b_{k,i}(v)$ and $g_i(v)$ represent systematic effects that are consistently observed across sessions for subject $i$. $\bfG(v)$ is the covariance matrix of the random effects vector $\bfb_{i}(v)$. It encodes population variance for each random effect, as well as the correlation between different random effects. For example, $Cor\left\{a_{1,i}(v),a_{4,i}(v)\right\}$ represents the correspondence between the direction and strength of subject $i$'s deviation from the population mean autocorrelation for tasks $1$ (emotion) and $4$ (relational), using the canonical HRF only. A strong positive correlation here would suggest that the same subjects tend to exhibit stronger or weaker autocorrelation, possibly due to their having a longer or shorter HRF than the canonical HRF, regardless of task.  $Cor\left\{a_{1,i}(v),b_{1,i}(v)\right\}$ represents the correspondence between the direction and strength of subject $i$'s deviation from the population mean on task $1$ and the effect of including the HRF derivative for the same task.  A strong anti-correlation here would suggest that including the HRF derivative has a bigger effect on subjects who exhibit stronger autocorrelation when using the canonical HRF -- in short, that inclusion of the HRF derivative achieves the goal of accounting for some population variability in HRF timing.

\subsubsection{Prewhitening Strategies}
\label{sec:prewhitening}

To evaluate the effectiveness of different prewhitening strategies on mitigating residual autocorrelation, we use an AR model with varying model order and varying degrees of regularization to estimate the prewhitening matrix $\bfW_v$. Specifically, we vary AR model order from $p=1$ to $p=6$. We also consider automatic selection of the optimal AR model order at each vertex using Akaike information criterion (AIC) \citep{sakamoto1986akaike} as proposed by \cite{luo2020improved}, with a maximum model order of $10$. The AR model coefficients and residual variance are estimated using the Yule-Walker equations \citep{Brockwell_Davis}. We consider both local and global regularization of the AR coefficients and white noise variance. Local regularization refers to surface smoothing with a 5mm FWHM Gaussian kernel; global regularization refers to smoothing with an infinitely-wide Gaussian kernel, nearly equivalent to averaging the AR coefficients across the cortex. In the case of optimal AR model order selection, we impute a value of $0$ for any AR coefficients above the selected model order prior to regularization.

Our R/C++ implementation of prewhitening is available in the open-source \texttt{BayesfMRI} R package \citep{mejia2022BayesfMRI}, which is compatible with cortical surface and ``grayordinates'' neuroimaging file formats via the \texttt{ciftiTools} R package \citep{pham2022ciftitools}. After estimating the prewhitening matrix $\bfW_v$ as described in Appendix \ref{app:PWalgo}, the response and design matrix are premultiplied at each vertex by $\bfW_v$, changing the GLM in (\ref{eqn:GLM}) to 
\begin{align}
    \tilde{\bfy}_v = \tilde{\bfX}_v\bfbeta_v + \tilde{\bfepsilon}_v, \quad \tilde{\bfepsilon}_v \sim \text{MVN}(\bfzero,\sigma^2\bfI), \label{eqn:pwGLM}
\end{align}
where $\tilde{\bfy}_v = \bfW_v\bfy_v$, $\tilde{\bfX}_v = \bfW_v\bfX_v$ and $\tilde{\bfepsilon}_v = \bfW_v\bfepsilon_v$. Note that the prewhitened design matrix $\tilde{\bfX}_v$ may vary across vertices when using a local approach to prewhitening. This increases the computational burden associated with GLM coefficient estimation: with a common design matrix $\tilde{\bfX}$, the model coefficients for all vertices can be estimated with a single matrix multiplication step as $(\tilde{\bfX}'\tilde{\bfX})^{-1}\tilde{\bfX}'\tilde{\bfY}$, where $\tilde{\bfY}=\left(\tilde{\bfy}_1,\dots,\tilde{\bfy}_V\right)$; when the design matrix varies spatially, however, we must perform $V$ matrix multiplications $(\tilde{\bfX}_v'\tilde{\bfX}_v)^{-1}\tilde{\bfX}_v'\tilde{\bfy}_v$ to obtain the coefficient estimate at each vertex $v$. Even more computationally burdensome is estimating $\bfW_v$ at each location, which involves performing $V$ different eigendecompositions. These obstacles are perhaps one reason that global prewhitening approaches are often preferred in a practical sense. To overcome these challenges, we have developed a highly computationally efficient implementation using parallelization and C++ backend code. This implementation typically completes in approximately 1 minute per scan for the task fMRI data we analyze here.

\subsubsection{Evaluation Metrics}

To evaluate the performance of each prewhitening strategy, we take two approaches. First, we directly assess the degree of residual autocorrelation still present in each task fMRI dataset after prewhitening.  Using on the fitted GLS residuals at each vertex, we compute the autocorrelation index (ACI) as in Section \ref{sec:ACI}. We also perform a Ljung-Box (LB) test at every vertex \citep{ljung1978measure} to identify vertices exhibiting statistically significant levels of residual autocorrelation after prewhitening. We use the \texttt{Box.test} function in the \texttt{stats} R package, version 4.2.0. As in \citep{corbin2018accurate}, we use the first 100 volumes of each session and consider up to 20 lags. We consider two approaches to determine the degrees of freedom (DOF) for the test: accounting for the intercept only, or accounting for the intercept and the AR model coefficients. We consider the intercept-only approach for maximum comparability with \cite{corbin2018accurate}.\footnote{Though \cite{corbin2018accurate} did not state explicitly how DOF was determined in their analysis, their implementation used the Matlab Ljung-Box test function, where ignoring the model DOF is the default. Further, the higher-order SPM FAST models considered in \cite{corbin2018accurate} contain more than 20 DOF, which would result in negative DOF for the LB test if taken into account.} When accounting for the AR($p$) model coefficients as well, the DOF for the LB test is $20 - \left[p*100/T\right] - 1$, where $T$ is the original number of volumes in the task fMRI session (given in Table \ref{tab:tasks}). We scale the number of AR coefficients $p$ by $100/T$ to account for the fact that the AR model parameters were estimated across the whole duration of the scan, not just the $100$ volumes used for the LB test. We determine vertices whose residuals exhibit significant autocorrelation based on those with $p<0.05$ after false discovery rate correction \citep{benjamini1995controlling}. We compute the proportion of vertices exhibiting significant autocorrelation before and after prewhitening with each technique to determine the ability of each prewhitening method to effectively eliminate autocorrelation.  

Second, we quantify false positives using resting-state fMRI data, assuming a false boxcar task paradigm.  For each resting-state fMRI dataset, we perform GLS using the estimated prewhitening matrix for each prewhitening strategy.  We then perform a t-test at every vertex. We correct for multiple comparisons across all vertices with Bonferroni correction to control the family-wise error rate (FWER) at $0.05$.  While Bonferroni correction is typically considered overly conservative for whole-brain voxel-wise analysis involving potentially hundreds of thousands of tests, here we are performing fewer than 6,000 tests per hemisphere. In previous work we have found Bonferroni correction to have similar power as permutation testing for the cortical surface resampled to a similar resolution \citep{spencer2022spatial}. We visualize the spatial distribution of false positive vertices and quantify the false positive rate and FWER before and after prewhitening. We obtain $95\%$ confidence intervals for the FWER using Agresti-Coull intervals for proportions (\cite{agresti1998approximate}).

\section{Results}
\label{sec:results}

\subsection{Overview}

We first examine the spatial patterns and factors influencing autocorrelation in task fMRI prior to any prewhitening, using a random effects analysis of task fMRI data from the HCP retest dataset. This allows us to understand to what degree residual autocorrelation varies across the cortex in task fMRI studies, which in turn helps inform an effective approach to prewhitening. We find that residual autocorrelation varies markedly across the cortex, and the spatial topology is influenced by the task being performed, the phase encoding direction, and systematic inter-subject differences. Effective modeling choices (e.g. HRF flexibility, nuisance regression) can mitigate autocorrelation, but their effects are relatively modest and they do not eliminate spatial variability. We then assess the ability of different AR-based prewhitening strategies to effectively mitigate residual autocorrelation and control false positive rates. We consider low- and high-order AR models, as well as optimal determination of the AR model order at each vertex. We also consider two opposing approaches to spatial regularization of AR model parameters: local spatial smoothing and global averaging across the cortex. We find that higher-order AR models that allow for spatial variability in AR model parameters are able to effectively mitigate autocorrelation, while global averaging and very low-order AR models retain substantial levels of autocorrelation. 

\subsection{Spatial patterns of residual autocorrelation}
%%% FIGURE 1 -- ACI BEFORE PREWHITENING
\begin{figure}
    \centering
    \includegraphics[width=0.9\textwidth]{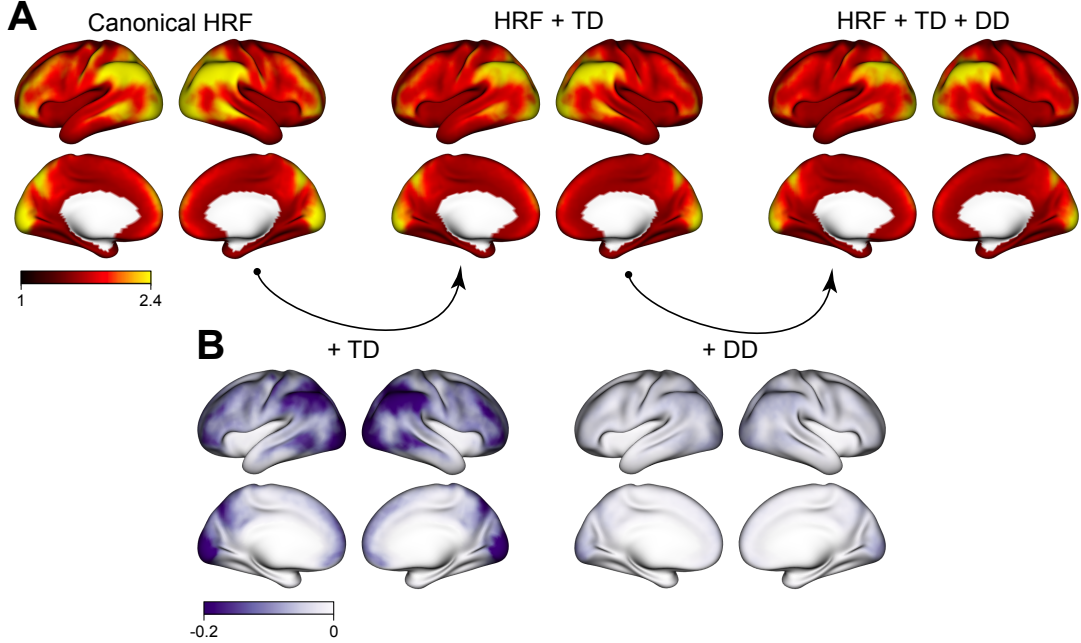}
    \caption{\small \textbf{Spatial patterns of autocorrelation and the effect of HRF modeling.} (A) The average autocorrelation index (ACI) across all subjects, sessions and tasks, for three different cases: assuming a canonical HRF, including the HRF temporal derivative (TD) to allow for differences in HRF onset; and including the HRF TD and dispersion derivative (DD) to allow for differences in HRF onset and duration.  (B) The reduction in ACI when HRF derivatives are included to allow for differences in HRF shape. Including the HRF TD has a sizeable effect in reducing autocorrelation; additionally including the HRF DD has a more subtle effect.\\}
    \label{fig:ACI_avg}
\end{figure}

\textbf{Figure \ref{fig:ACI_avg}} shows the autocorrelation index (ACI) across the cortex, averaged over all subjects, sessions, runs and tasks in the HCP retest study.  The spatial topology of autocorrelation across the cortex is striking: higher autocorrelation is seen in frontal, parietal and occipital areas, particularly the inferior parietal cortex and the occipital pole. Much lower autocorrelation is seen in the insula, the fusiform gyrus and the temporal pole, for example.  The regions with highest autocorrelation tend to be near the edge of the brain, possibly reflecting in part effects of motion.  Three different modeling strategies for the hemodynamic response function (HRF) are considered, ranging from rigid (canonical HRF) to more flexible (canonical HRF plus its temporal derivative (TD) and dispersion derivative (DD)). The degree of autocorrelation is reduced by the inclusion of HRF derivatives, illustrating that HRF mis-modeling is one source of autocorrelation that can be mitigated with more flexible HRF models. Panel B shows that inclusion of the TD serves to reduce autocorrelation most in the same areas where autocorrelation tends to be the highest, suggesting that the spatial patterns of autocorrelation may be due in part to heterogeneity in HRF onset and duration that varies across the cortex.  Even after including HRF derivatives, however, there are still marked differences in the degree of residual autocorrelation across the cortex. Including the dispersion derivative has a more subtle effect, compared with just including the temporal derivative. For the remaining analyses, we therefore focus on inclusion of the HRF temporal derivative only. 

\subsection{Random effects analysis of residual autocorrelation}

Here, we examine the sources of residual autocorrelation in task fMRI studies through a random effects analysis of repeated task fMRI scans from the HCP retest participants. We fit a series of general linear models (GLMs) to each task fMRI scan, varying the HRF modeling strategy and the number of motion regressors across GLMs. For each GLM, we quantify the autocorrelation index (ACI) of the model residuals at each surface vertex. We then fit the random effects model in (\ref{eqn:lmer_model}) at each vertex to quantify the contribution of task-specific differences, HRF modeling strategy, phase encoding direction, and number of nuisance regressors on the residual autocorrelation index. The inclusion of random effects accounts for systematic between-subject variability and allows us to understand population heterogeneity in these effects.

\begin{figure}
    \centering
    \includegraphics[width= 6in]{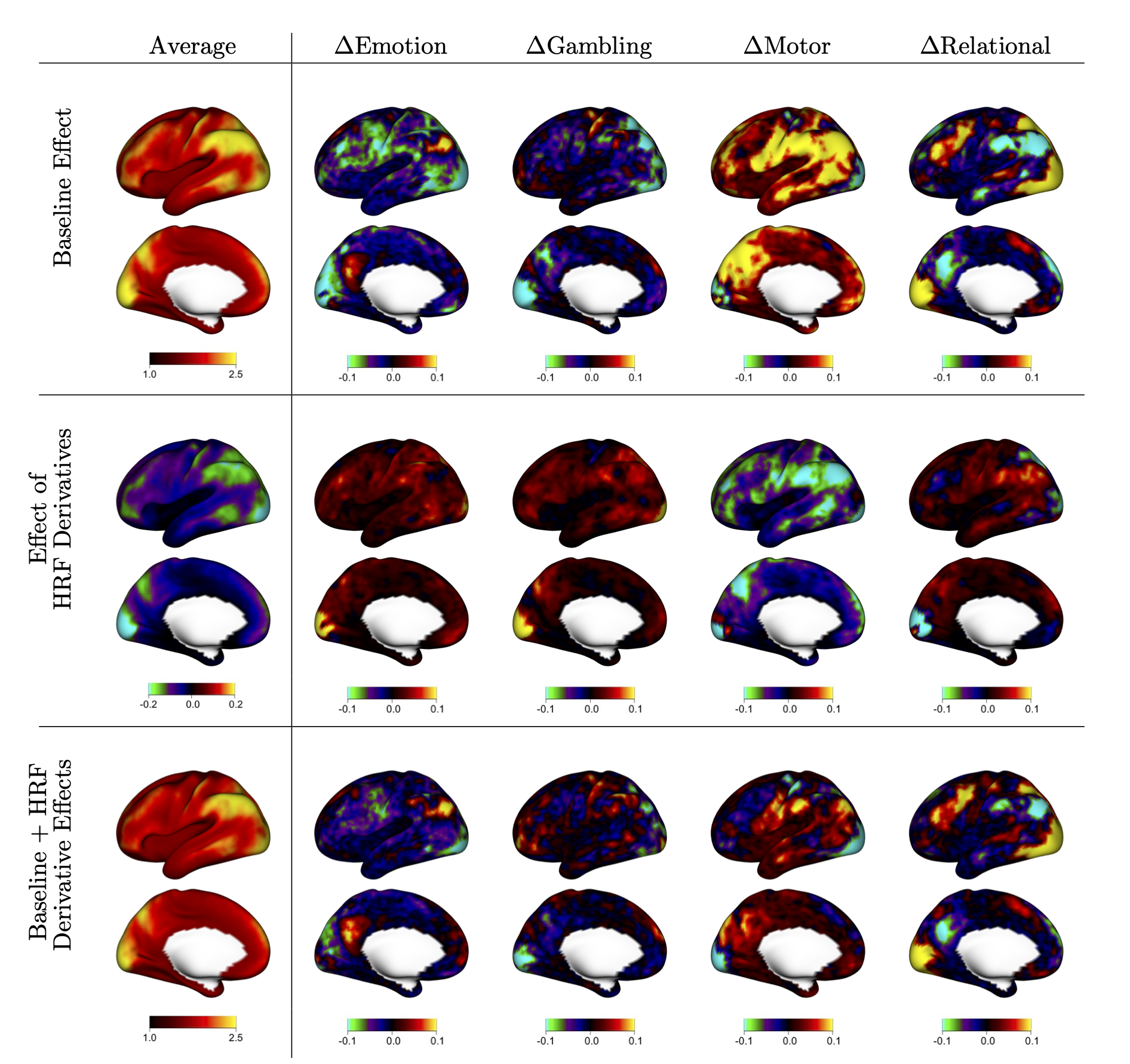}
    \caption{\small \textbf{Baseline autocorrelation index (ACI) and effect of including HRF derivatives by task,} based on the fixed effects (FEs) from the mixed effects model shown in equation (\ref{eqn:lmer_model}). The first column shows the average of FE estimates across tasks, indicating general spatial patterns of autocorrelation and the effect of HRF derivatives across the cortex. The other columns show the difference between each task and the average, to show areas of stronger or weaker effects during specific tasks. The first row shows the average ACI when assuming a canonical HRF ($\alpha_k(v)$); the second row shows the effect of including HRF derivatives to allow for heterogeneity in the shape of the HRF  ($\beta_k(v)$); the third row shows the sum of both effects, which represents the average ACI when including HRF derivatives in the model ($\alpha_k(v)+\beta_k(v)$).\\[20pt]
    }
    \label{fig:lmer_FE_task}
\end{figure}

\textbf{Figure \ref{fig:lmer_FE_task}} displays the fixed baseline effects associated with each task ($\alpha_{v}^{(k)}$, $k=1,\dots,4$ in model (\ref{eqn:lmer_model})), along with the effect of including the HRF derivative ($\beta_{v}^{(k)}$, $k=1,\dots,4$ in model (\ref{eqn:lmer_model})). The first column displays the mean effect over tasks $k=1,\dots,4$; the other columns display the difference between the task-specific effects and that mean effect. The baseline effects shown on the first row ($\alpha_k(v)$ in model \ref{eqn:lmer_model}) represent the average ACI when assuming a canonical HRF, including 12 motion regressors, and using the LR phase encoding direction. The average pattern is very similar to the mean ACI in the dataset shown in \textbf{Figure \ref{fig:ACI_avg}}.  The task-specific deviations show that ACI tends to be markedly higher or lower in certain regions depending on the task.  For example, the motor task tends to have higher residual ACI in many areas, whereas the emotion task tends to have lower residual ACI. These results show that there are systematic task-related effects on autocorrelation that vary across the cortex.  The HRF derivative effects shown on the second row represent the change in average ACI when including the HRF derivative in each GLM ($\beta_k(v)$ in model \ref{eqn:lmer_model}). The mean effect shows that including HRF derivatives tends to decrease ACI, particularly in areas where ACI tends to be the highest, as observed in \textbf{Figure \ref{fig:ACI_avg}}. The task-specific deviations show that more flexible HRF modeling has the strongest effect for the motor task, mimicing the more severe autocorrelation seen in the motor task. The areas most affected by flexible HRF modeling for each task tend to somewhat mimic the spatial patterns unique to each task, but do not fully account for them. The sum of both effects shown in the third row represent the average ACI when including HRF derivatives ($\alpha_k(v)+\beta_k(v)$ in model \ref{eqn:lmer_model}). The average image shows reduced autocorrelation on average, consistent with \ref{fig:ACI_avg}. The task-specific deviations show that task-related differences in residual autocorrelation are substantially reduced but not eliminated when using a more flexible HRF model.

\begin{figure}
    \centering
    \includegraphics[width= 6in]{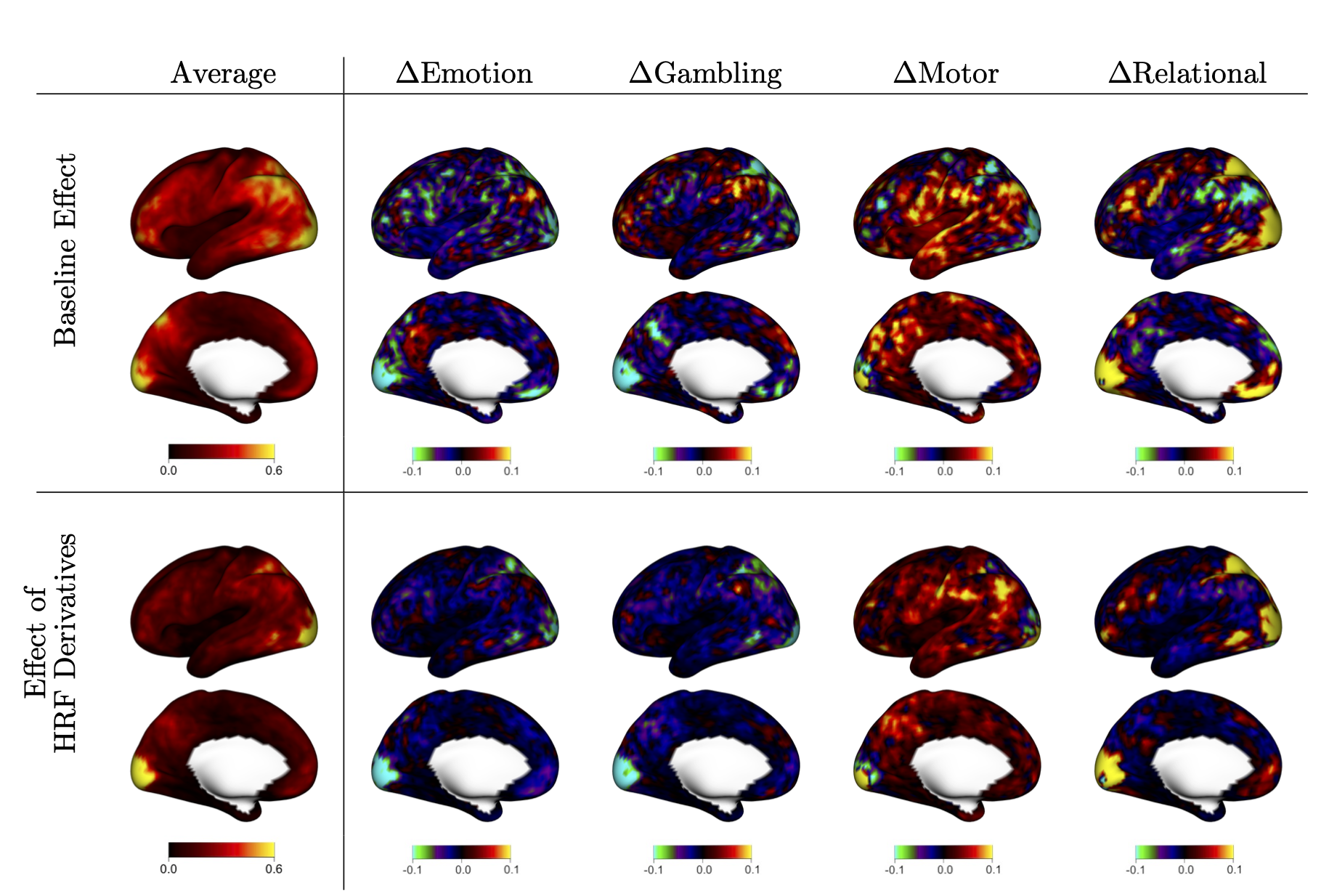}
    \caption{\small \textbf{Population variability in the effects shown in Figure \ref{fig:lmer_FE_task}}, based on the random effect (RE) standard deviations (SD) from model (\ref{eqn:lmer_model}). The first column shows the average across tasks, indicating general spatial patterns of population variability. The other columns show the difference between each task and the average, indicating areas of greater (warm colors) or lesser (cool colors) variability during specific tasks. The first row shows variability in autocorrelation when assuming a canonical HRF ($a_{k,i}(v)$); the second row shows variability in the effect of using HRF derivatives to allow for differences in HRF shape ($b_{k,i}(v)$).\\}
    \label{fig:lmer_RE_task}
\end{figure}

\textbf{Figure \ref{fig:lmer_RE_task}} displays the random effects associated with each fixed effect shown in \textbf{Figure \ref{fig:lmer_FE_task}}. These random effects represent reliable between-subject differences observed across repeated sessions from each subject.  The scale of the values is standard deviation, so they share the same units as the fixed effects. The figure organization is the same as that of \textbf{Figure \ref{fig:lmer_FE_task}}, with the first column representing the mean over tasks and the remaining columns representing differences between each task and that mean. The mean effects show that there is substantial population heterogeneity in the degree of residual autocorrelation, as well as in the reduction in residual autocorrelation achieved by more flexible HRF modeling. In general, the spatial patterns mimic that of the fixed effects: areas that tend to have higher autocorrelation on average also tend to exhibit greater systematic variability across subjects, and areas that benefit more from flexible HRF modeling on average also tend to exhibit the most population variability in the degree of reduction.  

% Figure 4 -- LMM RESULTS: EFFECTS OF RP24 
\begin{figure}
\centering
\begin{subfigure}{0.45\linewidth}
    \centering
    \fbox{\includegraphics[width=0.9\linewidth]{{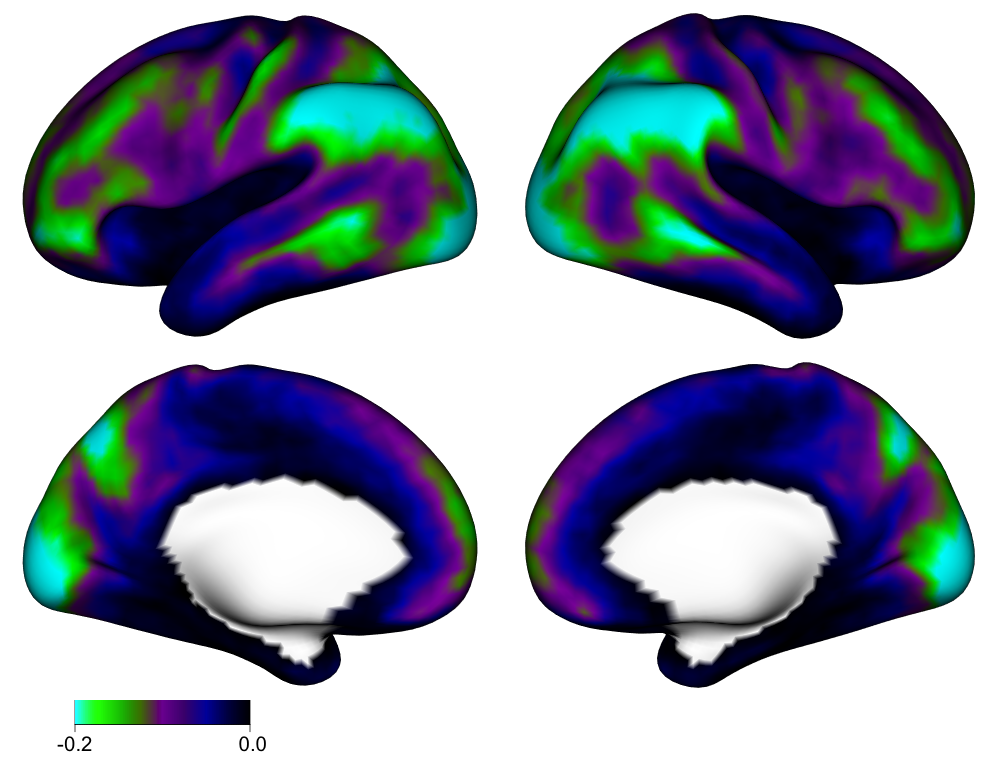}}}
    \caption{Fixed effects of RP24}
\end{subfigure}
\begin{subfigure}{0.45\linewidth}
    \centering
    \fbox{\includegraphics[width=0.9\linewidth]{{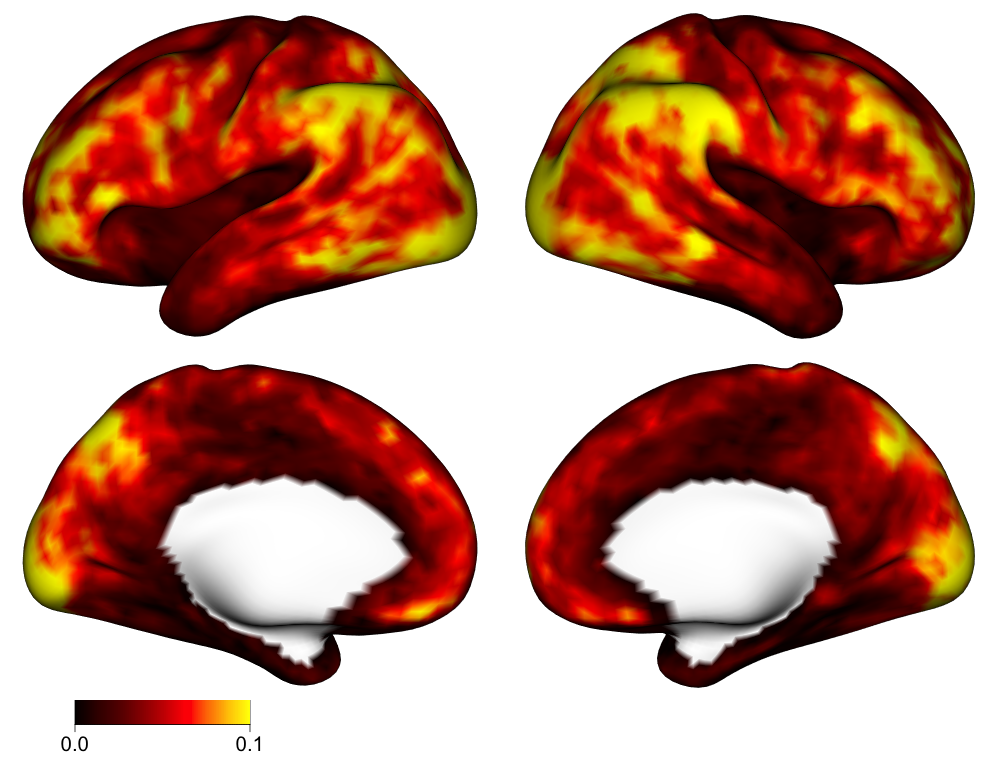}}}
    \caption{Random effects of RP24}
\end{subfigure}
\caption{\small \textbf{Effect of including additional motion regressors on autocorrelation index (ACI),} based on the fixed effects in model (\ref{eqn:lmer_model}). Values represent the decrease in ACI associated with including 24 rather than 12 motion regressors. Panel (A) shows the fixed effects at every vertex; panel (B) shows the random effect standard deviation at every vertex.}
    \label{fig:lmer_RP}
\end{figure}
    
\textbf{Figure \ref{fig:lmer_RP}} shows fixed and random effects of including additional motion regressors (24 versus 12) in the GLM on the degree of residual autocorrelation across the cortex. In the GLM with 12 motion regressors, the six realignment parameters (RPs) plus their one-back differences are included as covariates; in the model with 24 motion regressors, their square of each term is also included. Panel (A) shows that on average, including additional motion regressors decreases residual autocorrelation. This illustrates that without adequate nuisance signal modeling, temporally correlated noise such as that arising from head motion will be at least partly absorbed into the residuals, which will consequently exhibit greater autocorrelation. The spatial patterns mimic the spatial topology of baseline autocorrelation seen in Figures \ref{fig:ACI_avg} and \ref{fig:lmer_FE_task}, suggesting that thorough noise modeling helps to alleviate, without eliminating, the spatial heterogeneity in autocorrelation across the cortex. Panel (B) shows that there is population heterogeneity in the benefit of including additional motion regressors on autocorrelation, particularly in those areas with the most benefit on average.

% Figure 5 -- CORRELATION OF RANDOM EFFECTS
\begin{figure}
    \centering
    \includegraphics[page=2, width=0.65\textwidth, trim=-1in 0 0 0, clip]{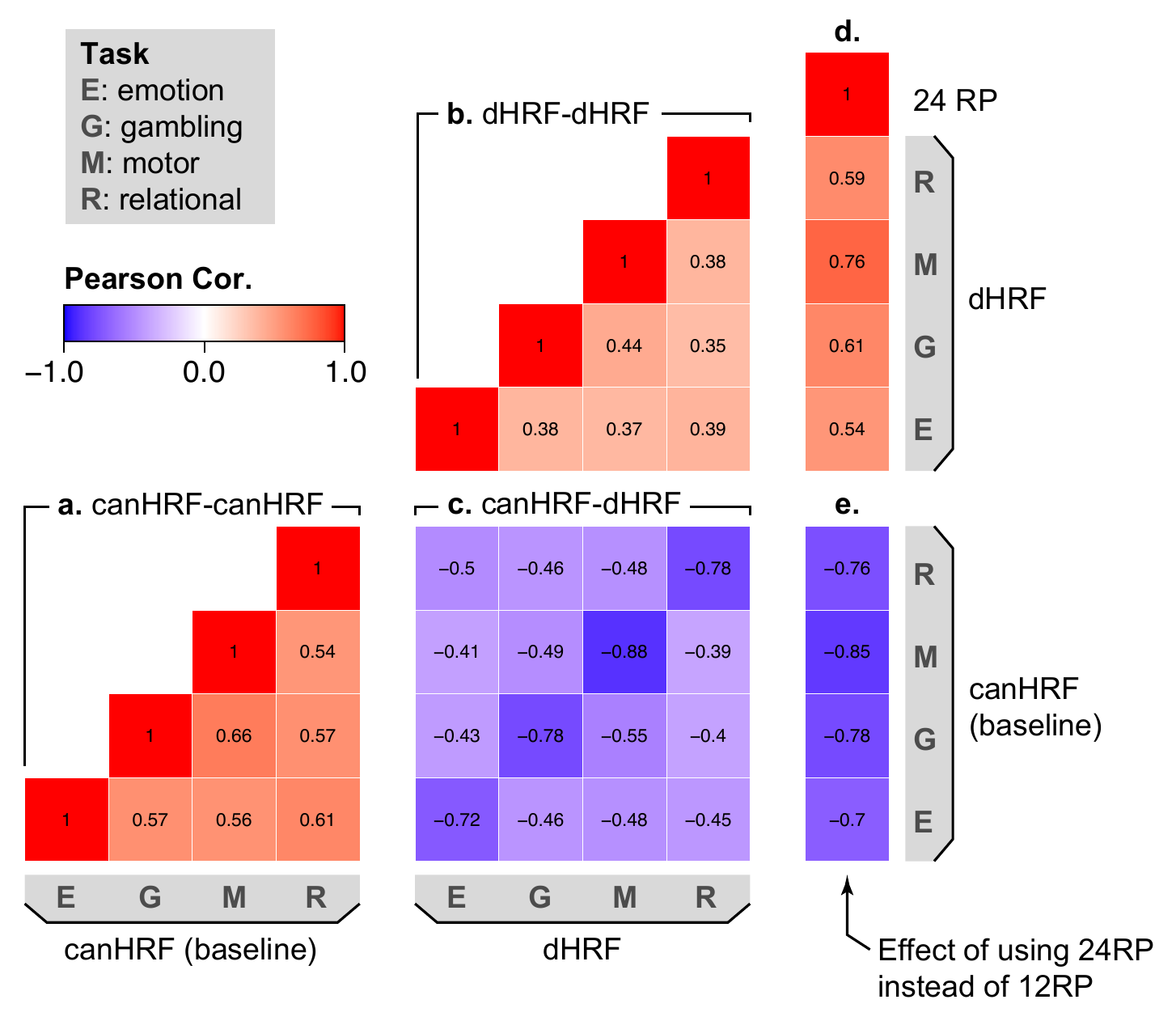}
    \caption{\small \textbf{Correlation among sources of population variability in autocorrelation,} based on the random effect (RE) correlations in model (\ref{eqn:lmer_model}). Values represent the mean across all vertices. Positive values (warm colors) between two effects indicate that subjects who exhibit higher values in one effect also tend to exhibit higher values in the other effect. Negative values (cool colors) indicate that subjects who exhibit higher values in one effect tend to exhibit lower values in the other effect.  The lower triangle of the correlation matrix is divided into several parts, representing the correlation between: a) different tasks in the baseline level of autocorrelation; b) different tasks in the effect of HRF derivatives on autocorrelation; c) the baseline level of autocorrelation and the effect of including HRF derivatives on autocorrelation; d) the effect of including additional motion regressors and the effect of including HRF derivatives; e) the effect of including additional motion regressors and the baseline level of autocorrelation.}
    \label{fig:RE_correlations}
\end{figure}

\textbf{Figure \ref{fig:RE_correlations}} takes a deeper look at population heterogeneity by examining the correlation between different random effects in the model. The lower triangle of the matrix is divided into several blocks, representing the different random components of the model. Blocks (A) and (B) show that there is moderate correlation between the different tasks in terms of the baseline effect and the effect of including HRF derivatives. This shows that for a particular subject, the spatial topology of autocorrelation and HRF shape are similar but not identical across different tasks. In block (C), the diagonal elements show a strong negative correlation between the baseline level of autocorrelation and the reduction in autocorrelation due to flexible HRF modeling. This shows that subjects having higher (or lower) baseline autocorrelation tend to benefit more (or less) from flexible HRF modeling. The strong but not perfect correlations suggest that accounting for between-subject differences in HRF shape reduces, but does not eliminate, systematic population variability in residual autocorrelation. Turning to the effect of motion, the strongly negative correlations in block (E) show that including additional motion regressors (24 instead of 12) reduces autocorrelation most for subjects who tend to exhibit stronger baseline autocorrelation. The moderate positive correlations in block (D) show that subjects who benefit more from more flexible HRF modeling also tend to benefit more from inclusion of additional motion regressors, suggesting that these two approaches may be complementary in reducing residual autocorrelation.

% Figure 6 -- LMM RESULTS: EFFECT OF RL  --- only dHRF related results
\begin{figure}
\begin{subfigure}{0.49\textwidth}
    \centering
    \includegraphics[height=6cm, trim=0 0 0 -2cm, clip]{{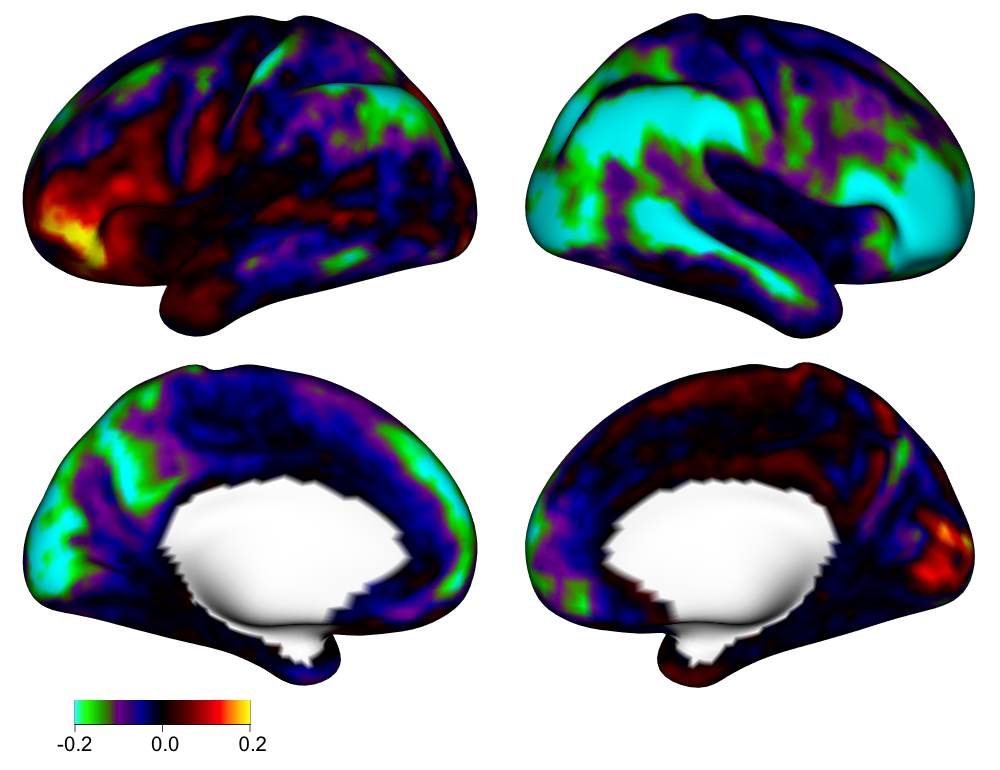}}
    \caption{Fixed effect of RL acquisition}
\end{subfigure}
\begin{subfigure}{0.49\textwidth}
    \centering
    \begin{tabular}{cc}
    Before Correction & After Correction \\
    \includegraphics[height=5.5cm, trim = 29cm 0 4cm 0, clip]{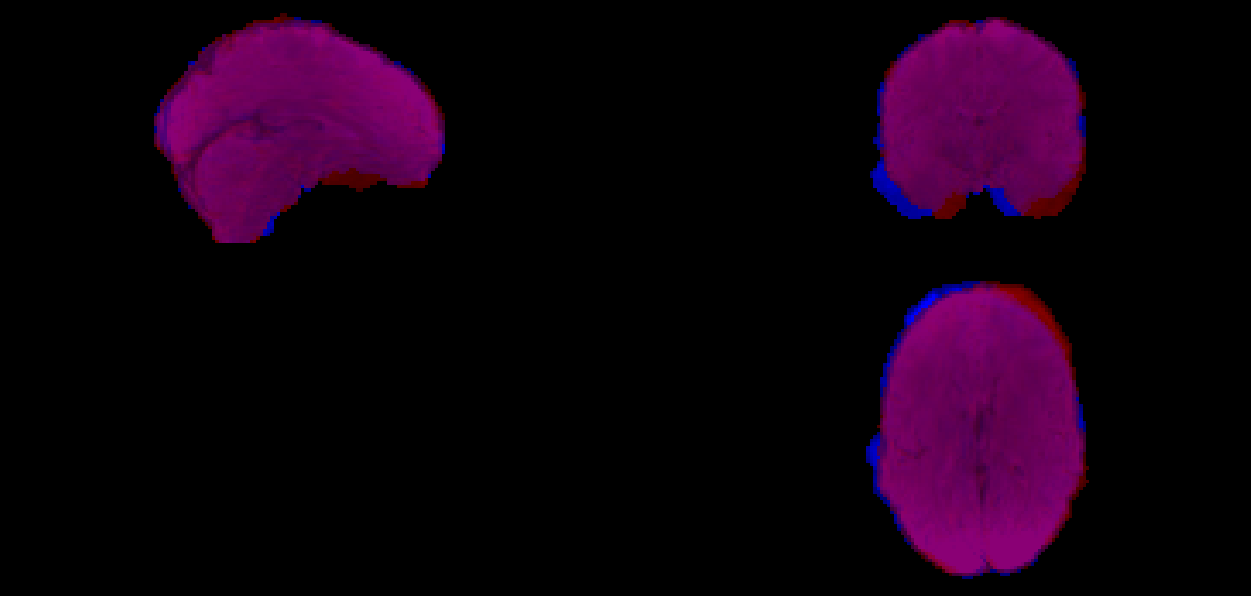} &
    \includegraphics[height=5.5cm, trim = 29cm 0 4cm 0, clip]{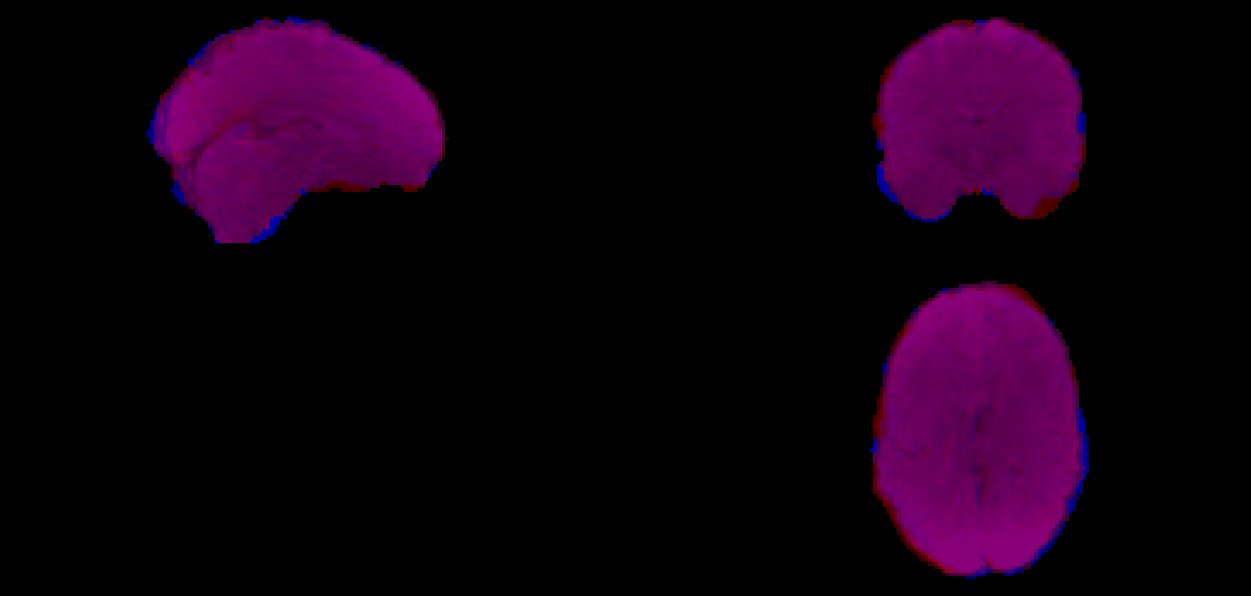}
    \end{tabular}
    \caption{LR (blue) and RL (red) distortions}
\end{subfigure}
\centering
\caption{\small \textbf{Effect of phase encoding direction on autocorrelation.} (A) Fixed effect of RL phase encoding direction in model (\ref{eqn:lmer_model}). Values represent the differences in average autocorrelation index (ACI) at each vertex when the RL (versus LR) phase encoding direction is used during image acquisition. Cool colors on the right lateral cortex, for example, indicate that RL acquisitions tend to have reduced autocorrelation in those areas compared with LR acquisitions. The effect of phase encoding direction is clearly lateralized, with the RL acquisition resulting in relatively lower autocorrelation on the right side of each hemisphere and higher autocorrelation on the left side of each hemisphere. This is likely due to distortions induced by the RL and LR phase encoding directions, even after distortion correction. (B) Mean rest fMRI image for a single subject (103818) for LR (blue) and RL (red) runs during the same session, before and after distortion correction, shown in neurological convention. Lateralized distortions persist after distortion correction, based on the imperfect overlap between the LR and RL runs.}
\label{fig:lmer_acq}
\end{figure}

% Table 2
\begin{table}[H]
    \caption{Settings for the simulation study shown in Figure \ref{fig:acq_sim}.}
    \centering
    \begin{tabular}{c|c|c|c}
        Tissue Class & Number of Voxels & AR(3) Model & ACI \\
        \hline
        White Matter & 11 voxels & $(0.1,\ 0.0,\ 0.0)$ & 1.1 \\
        Gray Matter & 2 voxels & $(0.425,\ 0.25,\ 0.1)$ & 2.3\\
        CSF & 3 voxels & $(0.5,\ 0.3,\ 0.1)$ & 4.5 \\
        Background & 11 voxels & white noise only & 1 \\
    \end{tabular}
    \label{tab:acq_sim}
\end{table}

\textbf{Figure \ref{fig:lmer_acq}A} displays the fixed effects associated with the utilizing an RL phase encoding direction acquisition, compared with an LR phase encoding direction. The effect of phase encoding direction on residual autocorrelation is clearly lateralized, with the RL phase encoding direction generally producing less autocorrelation on the right side of each hemisphere (the lateral cortex of the right hemisphere, and the medial cortex of the left hemisphere). This is likely due to lateralized distortions induced by the RL and LR phase encoding directions even after distortion correction, as shown in \textbf{Figure \ref{fig:lmer_acq}B}.  The spatial distribution of residual autocorrelation is therefore sensitive to specific acquisition. This may result in an increased risk of false positives in certain areas, depending on the acquisition method. For example, using an LR phase encoding direction, residual autocorrelation is generally higher within the right lateral cortex. This will result in higher rates of false positives compared with the left lateral cortex if not accounted for with prewhitening techniques that account for such spatial discrepancies in residual autocorrelation. 

% Figure 7 --- Simulation Results
\begin{figure}
    \centering
    \begin{subfigure}{0.30\linewidth}
    \centering
        \includegraphics[angle=90,origin=c, width=1\textwidth, trim=1in 0 0 0, clip]{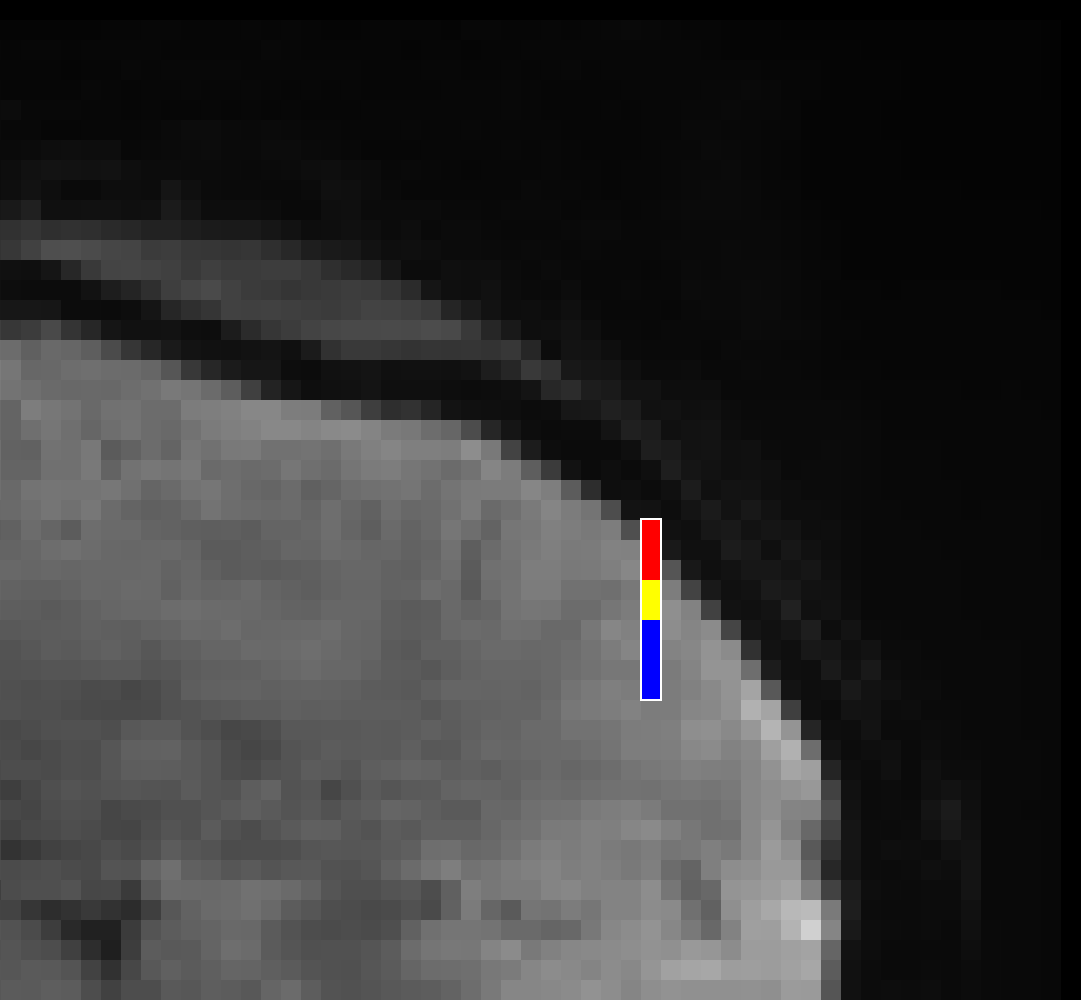}
        \caption{\small Voxels used in the simulation}
    \end{subfigure}
    \begin{subfigure}{0.69\linewidth}
    \centering
        \begin{tabular}{c|c}
        %  \multicolumn{6}{c}{} \\
        % & {True} & &  {After} & {After} & \\
        % & {ACI} &  & {distortion} & {correction} & {Bias}\\
        % \includegraphics[height=2in, trim=1.5in 1in 0.15in 0, clip]{RL_LR/F_bias_avgACI_tissue.pdf} \\
         & \includegraphics[width=2.1in, trim=0in 2.1in 0in 0, clip]{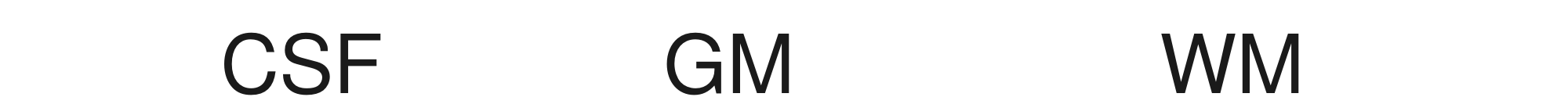} \\
        \hline
        {True ACI} & \includegraphics[width=2in, trim=0in 1.4in 0in 0.1in, clip]{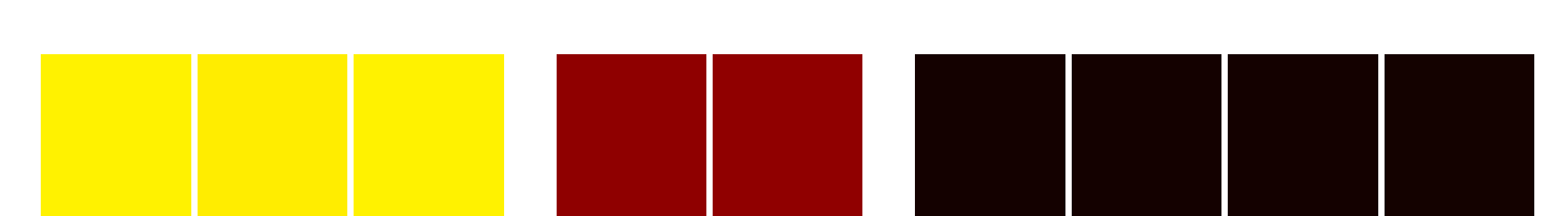} \\
        {After distortion} & \includegraphics[width=2in, trim=0in 1.4in 0in .1in, clip]{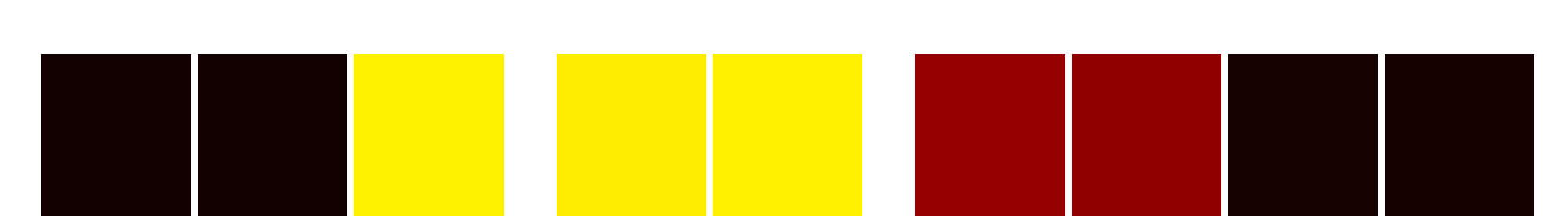} \\ 
        {After correction} & \includegraphics[width=2in, trim=0in 1.4in 0in .1in, clip]{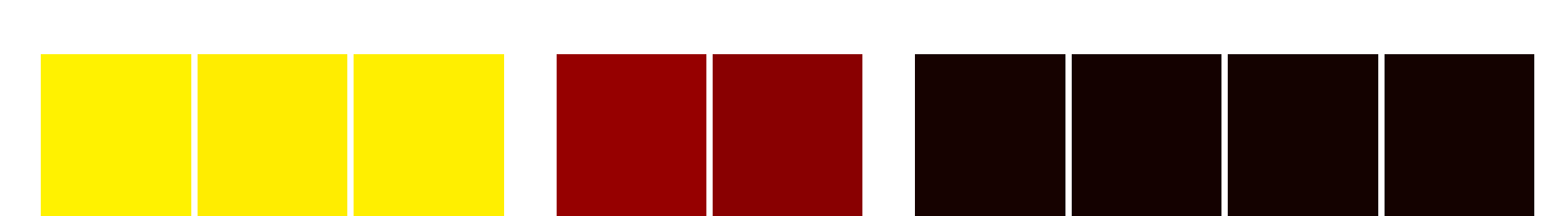} \\
         & \includegraphics[width=1.2in,  trim=.5in 0.2in 0.5in 0.3in, clip]{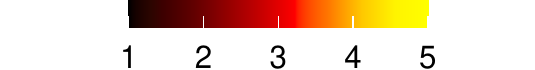} \\
        & \\
        {Bias} & \includegraphics[width=2in, trim=0in 1.4in 0in .4in, clip]{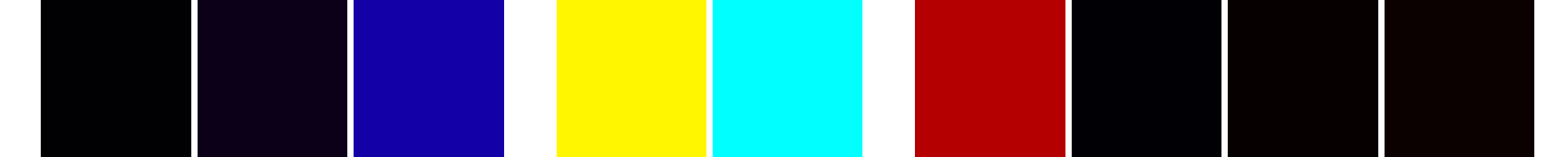} \\
         & \includegraphics[width=1.2in, trim=0.5in 0.2in 0.5in 0.3in, clip]{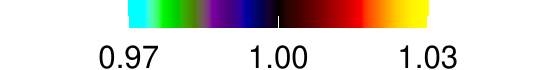} \\
         \hline
    \end{tabular}
    \centering
    \caption{\small ACI and bias in the simulation study\\[10pt]}
\end{subfigure}

\caption{\small \textbf{Effect of distortions on the spatial topology of autocorrelation}. (A) A nine-voxel sequence contains three voxels from cerebral spinal fluid (CSF, red), two voxels from gray matter (GM, yellow), and four voxels from white matter (WM, blue). These were padded by 26 background voxels on the left and 15 additional WM voxels on the right to avoid edge effects. (B) An AR(3) was used to generate autocorrelated timeseries within each tissue class, resulting in the true autocorrelation indices (ACI) shown on the top row. The ACI of the timeseries after after forward-direction distortion are shown on the second row and after distortion correction on the third row. While distortion correction clearly helps to resolve changes in ACI induced by the distortions, the fourth row shows that there is still bias (after/true) present after correction. Namely, the GM voxel neighboring CSF has increased ACI, and the GM voxel neighboring WM has decreased ACI. There is also a lesser amount of bias in the CSF and WM voxels neighboring GM.}
\label{fig:acq_sim}
\end{figure}

We examine this further through a small simulation study from HCP subject $103818$, shown in \textbf{Figure \ref{fig:acq_sim}}.  We consider a strip of nine voxels overlapping with the edge of the brain of the subject, shown in \textbf{Figure \ref{fig:acq_sim}A}. These include, sequentially, three voxels in CSF (red), two cortical gray matter voxels (yellow), and four white matter voxels. In addition, we include 26 background voxels on the left and 14 additional WM voxels on the right in order to absorb any edge effects. We generate autocorrelated timeseries for each voxel using an AR(3) model with white noise variance equal to 1. The AR coefficients are chosen to induce low ACI in white matter, moderate ACI in gray matter, high ACI in CSF, and unit ACI (the minimum) in background voxels. \textbf{Table \ref{tab:acq_sim}} gives the AR coefficients and resulting ACI in each region. 

To examine the effect of the distortions introduced through lateralized phase encoding on the ACI, we estimate the distortion map from the original temporal mean undistorted brain images of the subject. We distort the simulated timeseries then apply distortion correction using the Anima image processing toolbox\footnote{https://github.com/Inria-Visages/Anima-Public}. \textbf{Figure \ref{fig:acq_sim}B} shows the true ACI in each voxel, the ACI after distortion, and the ACI after distortion correction based on the mean over 7000 randomly generated timeseries. The last row shows the bias between the ACI of the distortion-corrected data, as a proportion of the true ACI. We see inflated ACI in the gray matter voxel bordering CSF and diminished ACI in the gray matter voxel bordering white matter. This agrees with our findings in \textbf{\autoref{fig:lmer_acq}} and supports the hypothesis that the LR and RL acquisitions result in changes in autocorrelation in gray matter due to distortion-induced mixing of signals with white matter and CSF.

In sum, we observe marked spatial discrepancies in autocorrelation within cortical gray matter due to acquisition factors, modeling choices, task-related factors and individual differences. The following section evaluates the effect of different prewhitening strategies on mitigating autocorrelation, reducing spatial variability in autocorrelation, and controlling false positives.

\subsection{The effect of prewhitening strategy on autocorrelation and false positives}

Here, we apply several prewhitening strategies based on autoregressive (AR) modeling and evaluate their effect on both residual autocorrelation and false positive rates. Specifically, we consider AR model order ranging from 1 to 6, as well as a spatially varying ``optimal'' model order based on Akaike information criterion (AIC). For each AR model order, we also consider two spatial regularization levels of the AR model coefficient estimates: ``local'' regularization is achieved by surface-based spatial smoothing the coefficient estimates using a 5mm full width at half maximum (FWHM) Gaussian kernel \citep{pham2022ciftitools}; ``global'' regularization is achieved by averaging the estimates across all cortical vertices. For optimal model order selection, all remaining coefficients after the AIC-based model order $p^*$ (up to the maximum of 10) are set to zero prior to regularization. All of the prewhitening methods considered are implemented in the open-source \texttt{BayesfMRI} R package, version 2.0 \citep{mejia2022BayesfMRI, R}.

In this section, we include just 12 motion parameters (6 rigid body realignment parameters and their one-back differences) in each GLM by default. This is because when combined with effective prewhitening, there does not appear to be a benefit of including the squares of the motion parameters and their one-back differences. In fact, it appears to be slightly detrimental, as shown in \textbf{Figure \ref{fig:RP24_vs_RP12}}. This suggests that including these additional motion parameters no longer serves to reduce autocorrelation when effective prewhitening is performed, and the loss of degrees of freedom associated with the inclusion of superfluous covariates in the GLM actually worsens the performance of prewhitening. 

% Figure 8 -- OPTIMAL AR MODEL ORDER
\begin{figure}
\begin{subfigure}{1\linewidth}
    \centering
    \begin{tabular}{cccccc}
    \quad & Emotion & Gambling & Motor & Relational & \\[10pt]
    & \includegraphics[width=1.3in, trim=0 0 7in 0, clip]{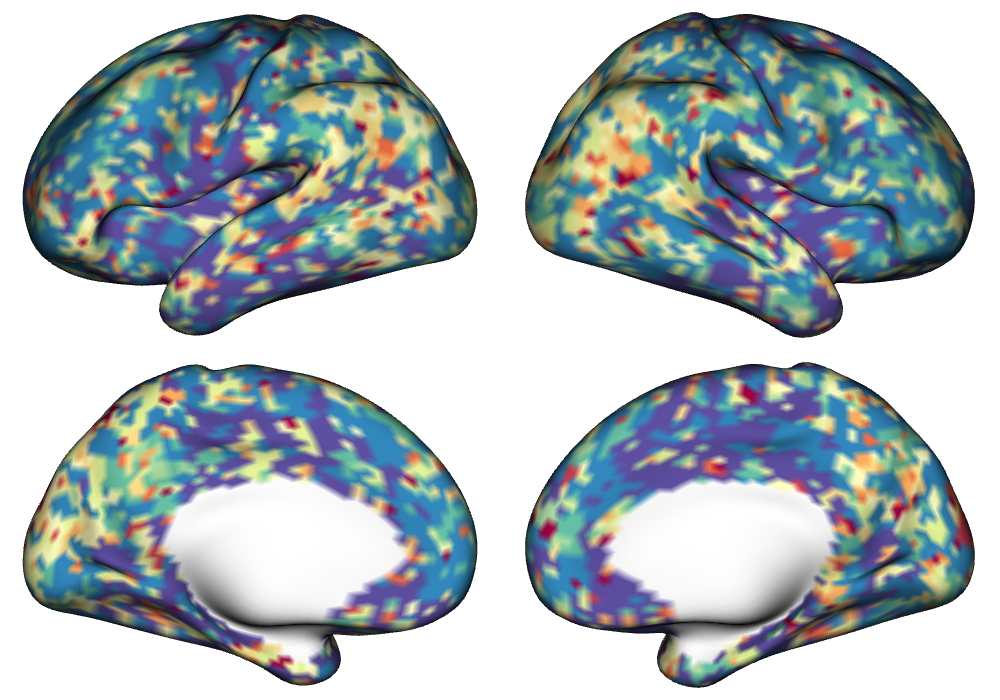} &
    \includegraphics[width=1.3in, trim=0 0 7in 0, clip]{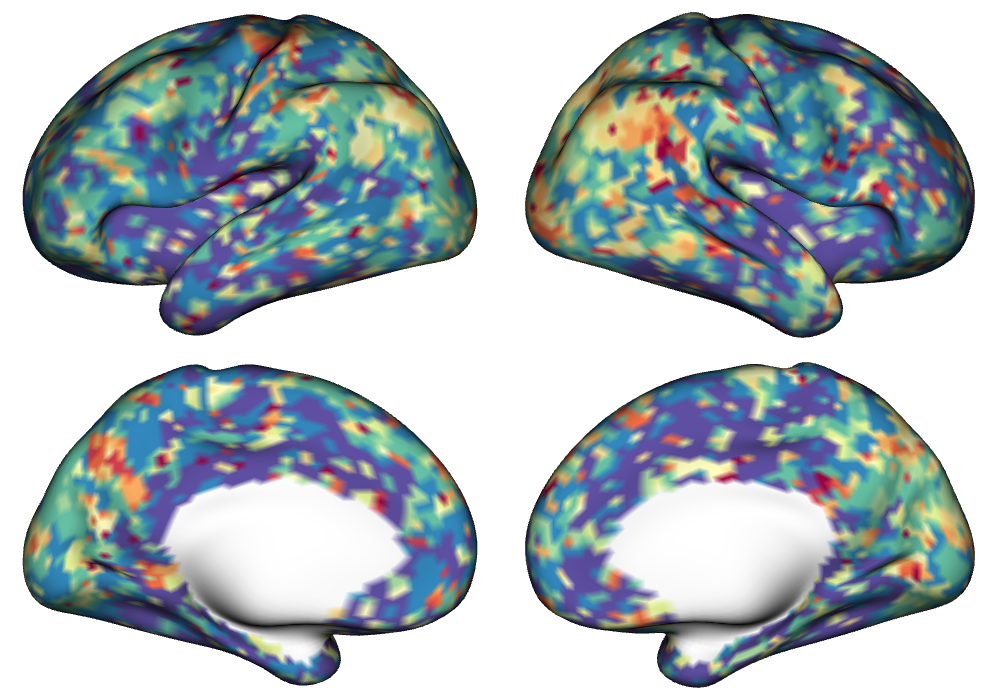} &
    \includegraphics[width=1.3in, trim=0 0 7in 0, clip]{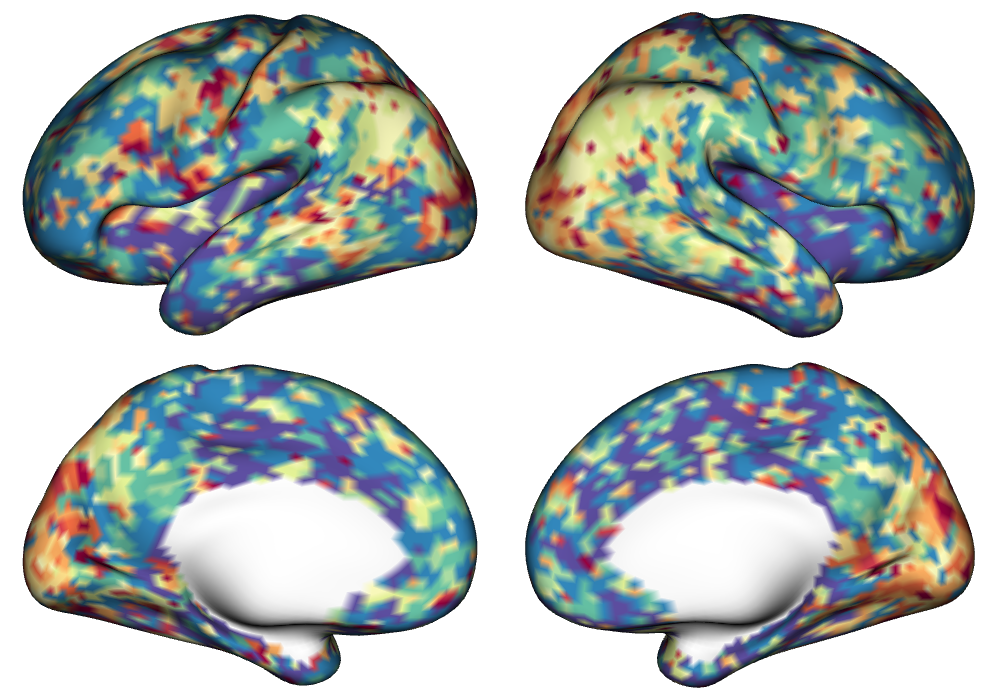} &
    \includegraphics[width=1.3in, trim=0 0 7in 0, clip]{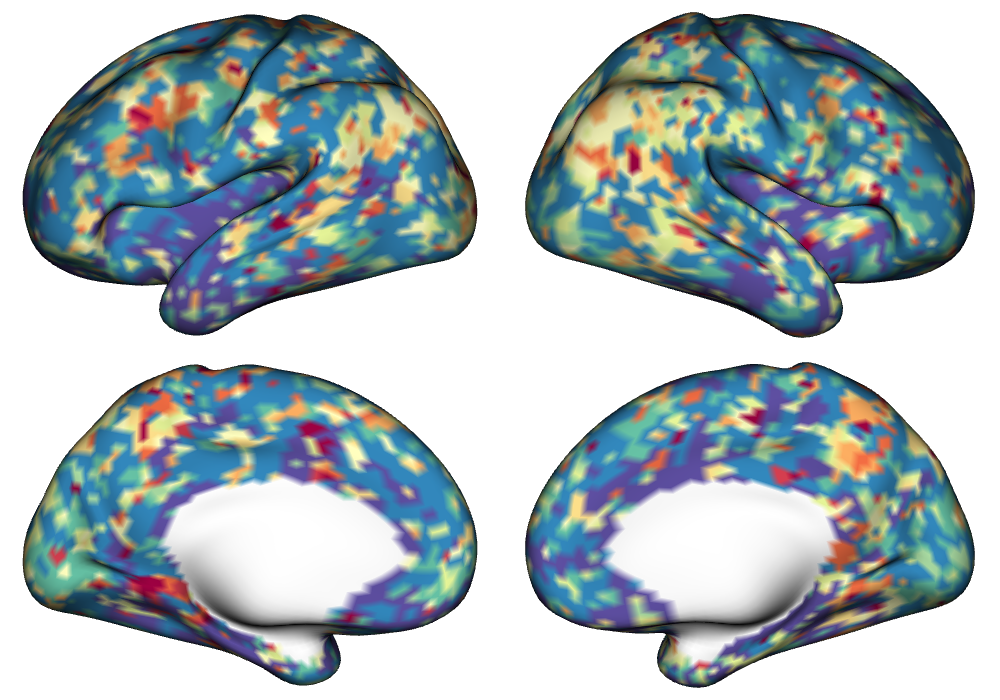} &
    \includegraphics[height=1.8in, trim=0 0 0 0, clip]{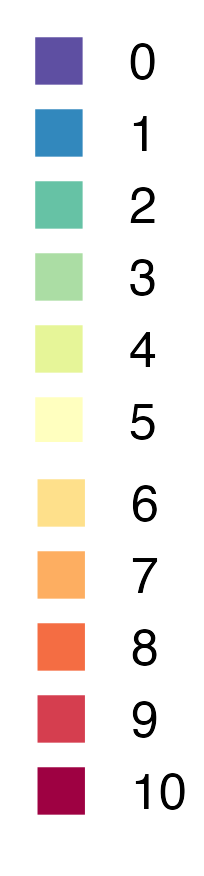}
    \end{tabular}
    \caption{Optimal AR model order for one subject for each task. \\[20pt]}
\end{subfigure}
\begin{subfigure}{1\linewidth}
    \centering
    \includegraphics[width=0.6\textwidth]{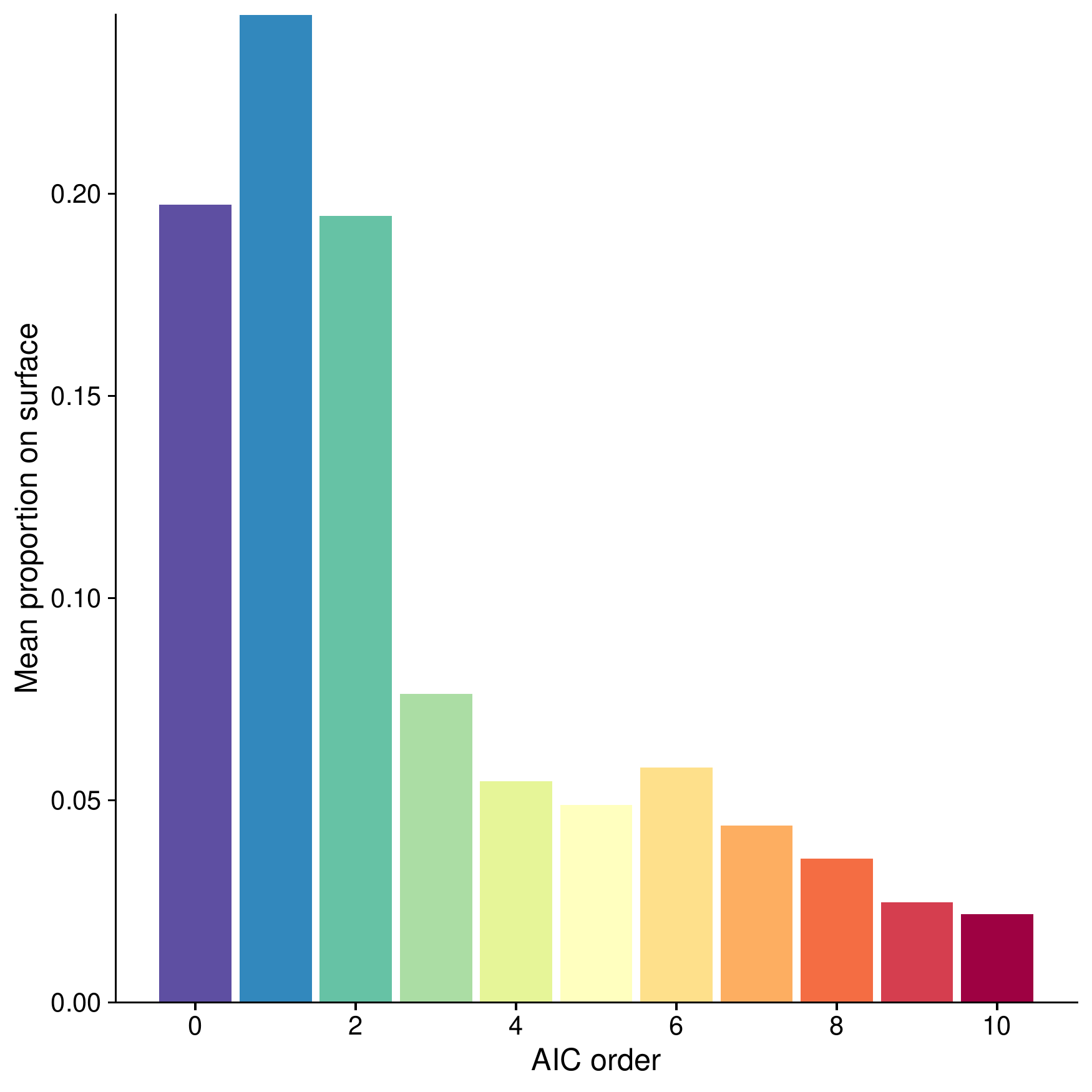}    
    \caption{Average distribution of optimal AR model order across the cortex.}
\end{subfigure}

\caption{\small \textbf{Optimal AR model order across the brain based on the Akaike information criterion (AIC).} (A) The optimal AR model order at every vertex for a single subject for each task. The optimal order clearly varies across the cortex and with the task being performed. (B) The distribution of optimal AR model order across all vertices, averaged over all subjects, sessions and tasks. The optimal AR model order is 2 or less for most vertices, but over $20\%$ of vertices have optimal AR model order of 3 to 6, while over $10\%$ have optimal order of 7 or higher.  }
\label{fig:optimal_AR_model_order}
\end{figure}

\textbf{Figure \ref{fig:optimal_AR_model_order}} displays the distribution of AIC-based AR model order, $p^*$, across the cortex.  Panel (A) shows the value of $p^*$ for a single run from a single subject across all four tasks.  The spatial patterns mimic the general patterns of autocorrelation strength seen in \textbf{Figure \ref{fig:ACI_avg}}.  Areas of higher autocorrelation generally require a higher AR model order (e.g. the inferior parietal cortex and the occipital pole), while areas of lower autocorrelation (e.g. the insula and the temporal pole) generally require a lower AR model order or no prewhitening at all ($p^*=0$). Differences across tasks can also been seen: for example, the motor task generally requires a higher AR model order, reflecting the stronger residual autocorrelation associated with the motor task as seen in \textbf{Figure \ref{fig:lmer_FE_task}}. The AIC-based model order is somewhat noisy, suggesting that some degree of regularization is needed. Panel (B) shows a histogram of AIC-based model order across all vertices. The proportions are averaged across all subjects, sessions and tasks. We see that on average, while the optimal AR model order is 2 or less for most vertices, over 30\% of vertices have optimal AR model order of 3 or higher, while over 10\% require an AR model order of 7 or higher. This again underscores the important spatial differences in residual autocorrelation across the cortex and the need for prewhitening methods that account for those differences to avoid under- or over-whitening in a given area.

\begin{figure}
    \centering
        \includegraphics[width= 6in]{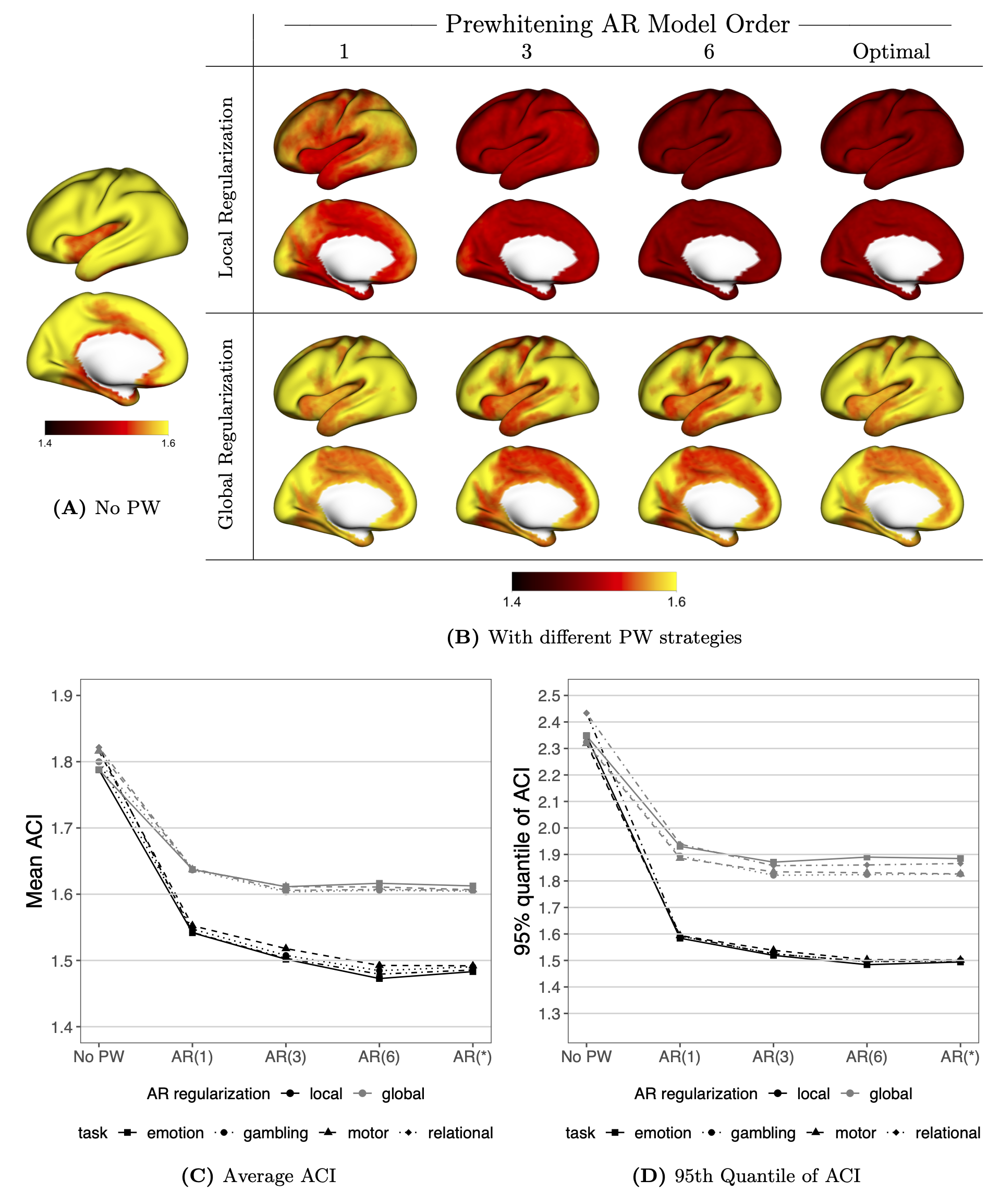}
    \caption{\small \textbf{The effect of prewhitening on autocorrelation index (ACI).} (A) Mean ACI over all subjects, sessions and tasks before prewhitening. (B) Mean ACI after prewhitening. Eight different prewhitening strategies are shown, based on four different AR model orders (1, 3, 6 and optimal at each vertex) and two different regularization strategies for AR model coefficients (local smoothing versus global averaging). Higher AR model order and allowing AR model coefficients to vary spatially results in substantially greater reduction in ACI. (C) Mean ACI over subjects and sessions, averaged across all vertices, by task and prewhitening method. Notably, allowing AR model coefficients to spatially vary reduces ACI much more than increasing AR model order.}
    \label{fig:prewhitening_ACI}
\end{figure}

\textbf{Figure \ref{fig:prewhitening_ACI}} shows the effect of each prewhitening strategy on the degree of residual autocorrelation.  Panel (A) displays the autocorrelation index (ACI) at each vertex averaged over all subjects, sessions and tasks, prior to any prewhitening.  Panel (B) displays the average ACI at each vertex after prewhitening with each strategy (varying AR model order, local versus global regularization of the AR model coefficients). Panel (C) displays the mean and 95th quantile of ACI across the cortex by task, averaged over all subjects and sessions, after prewhitening with each strategy. Panels (B) and (C) show that there is a dramatic difference between local and global regularization in terms of reducing autocorrelation: local regularization reduces ACI more and mostly eliminates the spatial variability in ACI. The combination of higher AR model order (e.g. AR(6)) with local regularization is the most effective at reducing ACI. Notably, the use of higher model orders in combination with global regularization is not very effective at reducing autocorrelation in many areas of the cortex: even an AR(1) model with coefficients that are allowed to spatially vary appears to be more effective than an AR(6) model with global regularization. Interestingly, the use of AIC to select the optimal model order at each vertex does not appear to be advantageous over fitting an AR(6) model at every vertex. It is worth noting that an AR(6) model encompasses lower-order AR models, since the higher coefficients can equal zero. Local regularization of the AR model coefficients may have the effect of shrinking those higher coefficients closer to zero when that is appropriate. Therefore, fitting an AR(6) model at each vertex, combined with local regularization, may allow for less aggressive prewhitening in those areas that exhibit less autocorrelation.

\begin{figure}
    \centering
        \includegraphics[width= 6in]{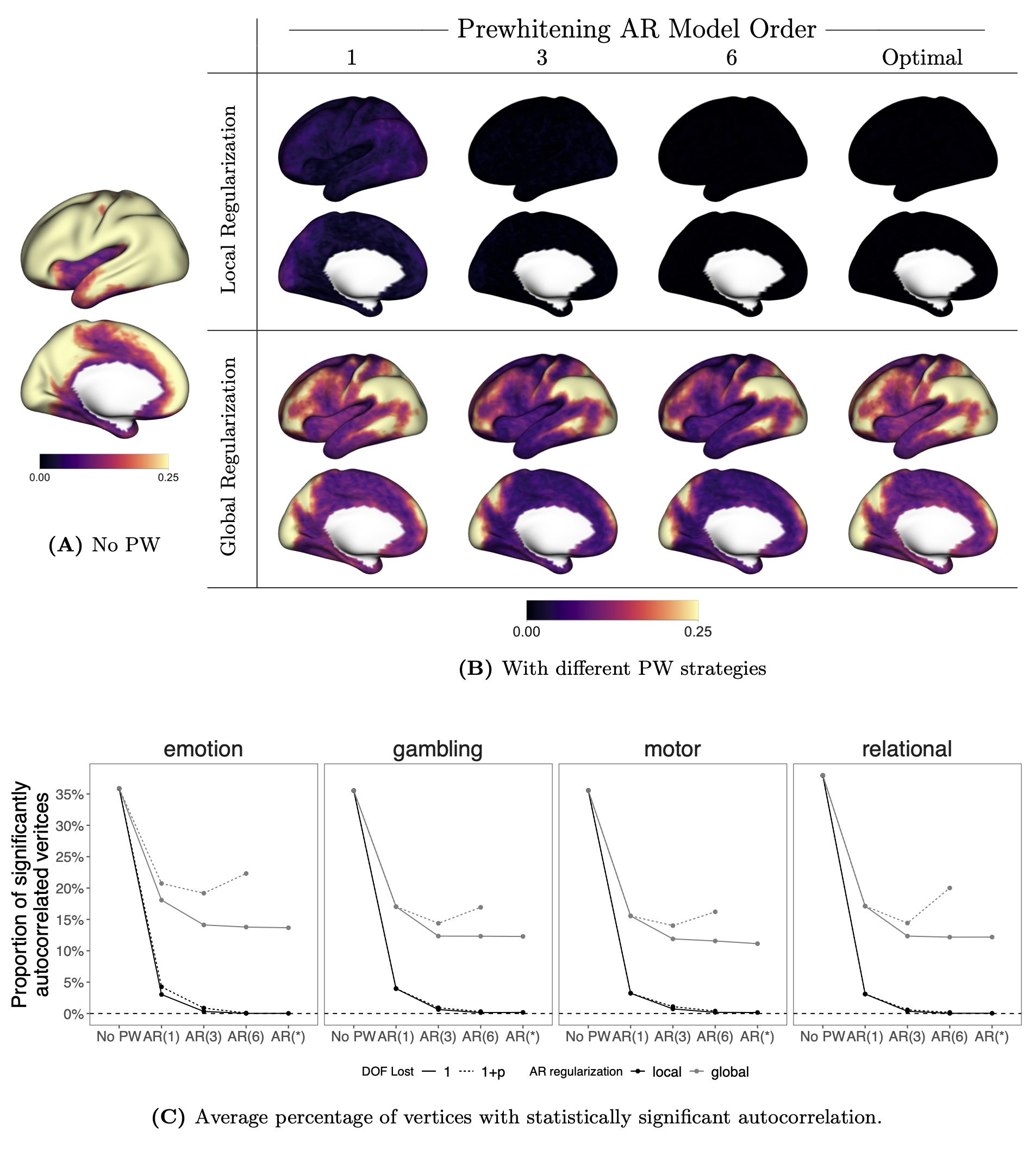}
     \caption{\small \textbf{The effect of prewhitening on the number of vertices with statistically significant autocorrelation.} (A) Proportion of sessions exhibiting significant autocorrelation at each vertex before prewhitening. (B) Proportion of sessions exhibiting statistically significant autocorrelation after prewhitening. Eight different prewhitening strategies are shown, based on four different AR model orders (1, 3, 6 and optimal at each vertex) and two different regularization strategies for AR model coefficients (local smoothing versus global averaging). Higher AR model order and allowing AR model coefficients to vary spatially results in substantially greater reduction in the number of vertices with statistically significant autocorrelation. Notably, allowing AR model coefficients to spatially vary has a greater effect than increasing AR model order. (C) Percentage of vertices with statistically significant autocorrelation, averaged across all subjects, sessions and tasks. Dotted lines correspond to accounting for the degrees of freedom (DOF) lost when estimating AR coefficients. Adopting an AR model order of 3 or higher and allowing AR coefficients to vary spatially results in virtually no vertices with statistically significant autocorrelation. }
\label{fig:prewhitening_LJB}
\end{figure}

\textbf{Figure \ref{fig:prewhitening_LJB}} displays the effect of prewhitening on the rate of vertices with statistically significant autocorrelation, based on performing a Ljung-Box (LB) test at every vertex \citep{ljung1978measure}. We correct for multiple comparisons by controlling the false discovery rate (FDR) at 0.05 using the Benjamini-Hochberg procedure \citep{benjamini1995controlling}. In panels (A) and (B), the value at each vertex represents the proportion of sessions that show significant autocorrelation across all subjects, sessions and tasks.  Panel (C) shows the proportion of significantly autocorrelated vertices in each session by task, averaged over all subjects and sessions. For now we will focus on the solid lines, which represent the results of the LB test when we assume a single degree of freedom lost in all models. The patterns in panels (A) and (B) mimic those seen in \textbf{Figure \ref{fig:prewhitening_ACI}}: local regularization of AR prewhitening parameters is much more effective at reducing autocorrelation than global regularization, and even a parsimonious (e.g. AR(1)) AR model with local coefficient regularization is more effective than a high-order AR model with global regularization. Panel (C) shows that AR-based prewhitening with local regularization (the black lines) essentially eliminates statistically significant autocorrelation in all vertices, particularly when using an AR model order of 3 or higher. A globally regularized, high-order AR model approach is less effective, reducing the proportion of significantly autocorrelated vertices to $10$-$15\%$. Note that this is similar to the performance of the optimal $12$-component SPM FAST model for data with TR=0.7 \citep{corbin2018accurate}. The much greater reduction in autocorrelation with local regularization illustrates the need to consider spatial differences in autocorrelation for effective prewhitening.

Note that in panel (C) of \textbf{Figure \ref{fig:prewhitening_LJB}}, we also consider the effect on the test result of accounting for the degrees of freedom (DOF) lost through the AR model fit ($1+p$) or just the intercept ($1$).\footnote{Note that for the optimal AR model order approach (AR(*)), we do not consider accounting for the DOF lost through model fitting when summarizing across vertices, since the DOF varies across vertices.} Though accounting for the AR fit in the total DOF is recommended, we account for the intercept only in (A), (B) and the solid lines in (C). This is done in order to replicate the analysis of \cite{corbin2018accurate}, which was based on the Matlab implementation of the Ljung-Box test, where ignoring the model DOF is the default. For more complex models involving many parameters, accounting for the model DOF generally results in apparently higher rates of autocorrelation, as seen in the U-shaped gray dashed lines in panel (c). This somewhat counterintuitive effect is simply a consequence of the loss in total degrees of freedom going from a more parsimonious model (e.g. AR(3)) to a more complex one involving more parameter estimates (e.g., AR(6)). Accounting properly for the DOF lost helps to avoid overestimating the performance of more highly parameterized models, which run the risk of overfitting to the data.

% Figure 11 -- FALSE POSITIVE RATES
\begin{figure}
\begin{subfigure}{1\textwidth}
    \centering
    \begin{tabular}{ccc}
       No   & AR(6), Local & AR(6), Global \\
       Prewhitening & Regularization &  Regularization \\
       \hline
    \includegraphics[height=2in, trim=0 0 7in 2cm, clip]{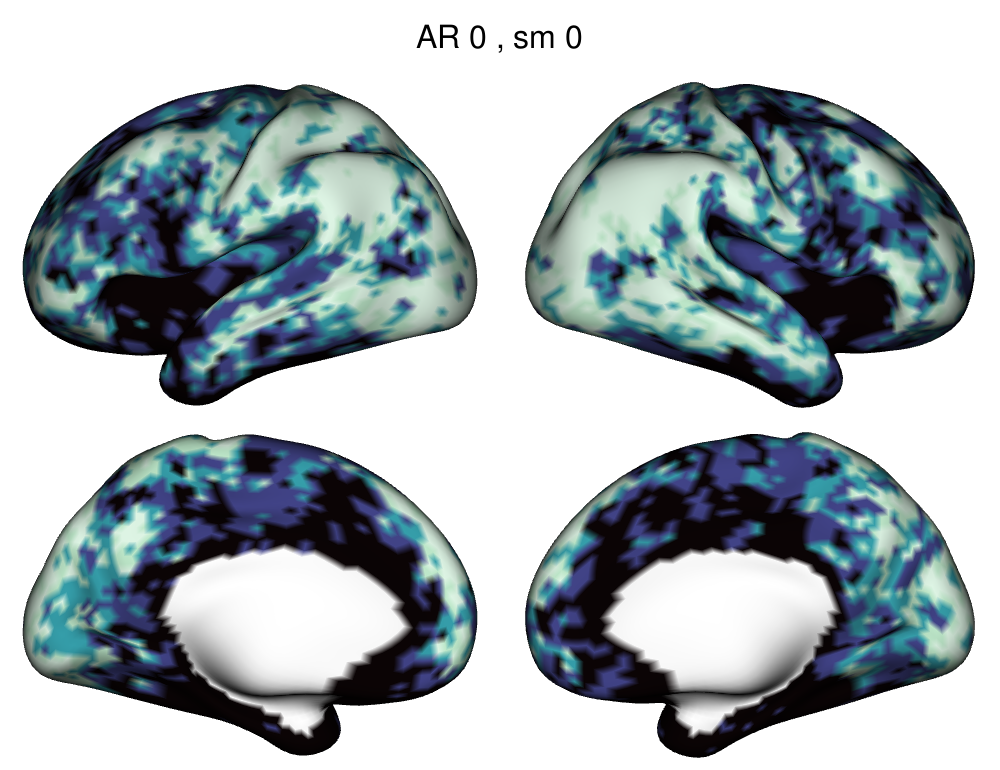} &
    \includegraphics[height=2in, trim=0 0 7in 2cm, clip]{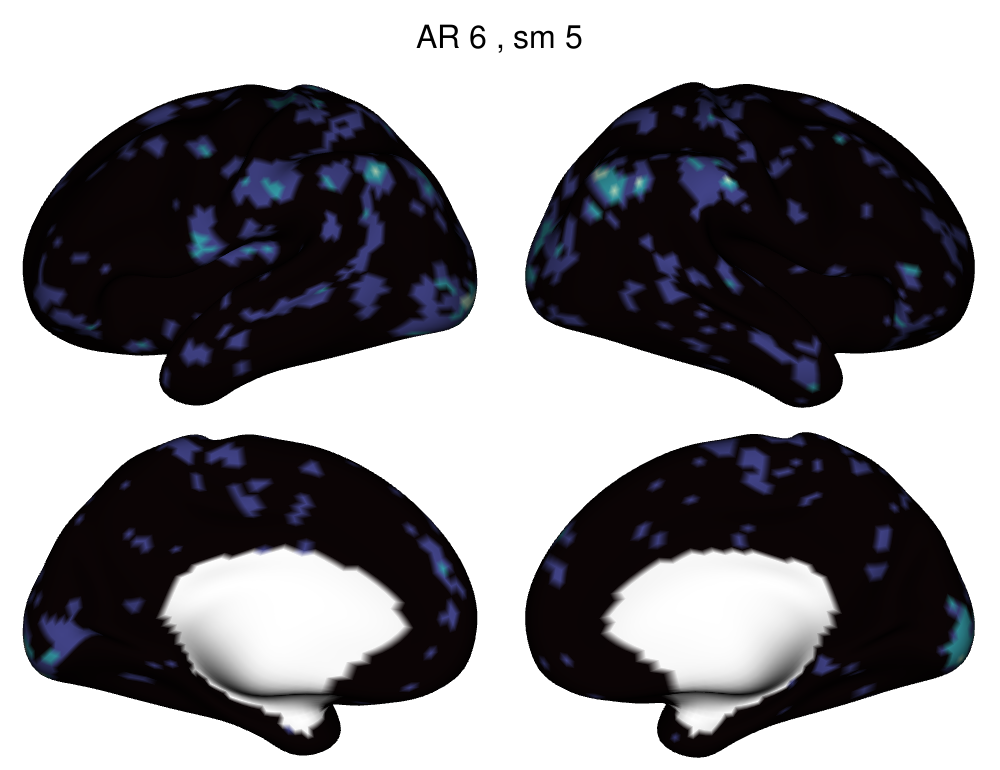} &
    \includegraphics[height=2in, trim=0 0 7in 2cm, clip]{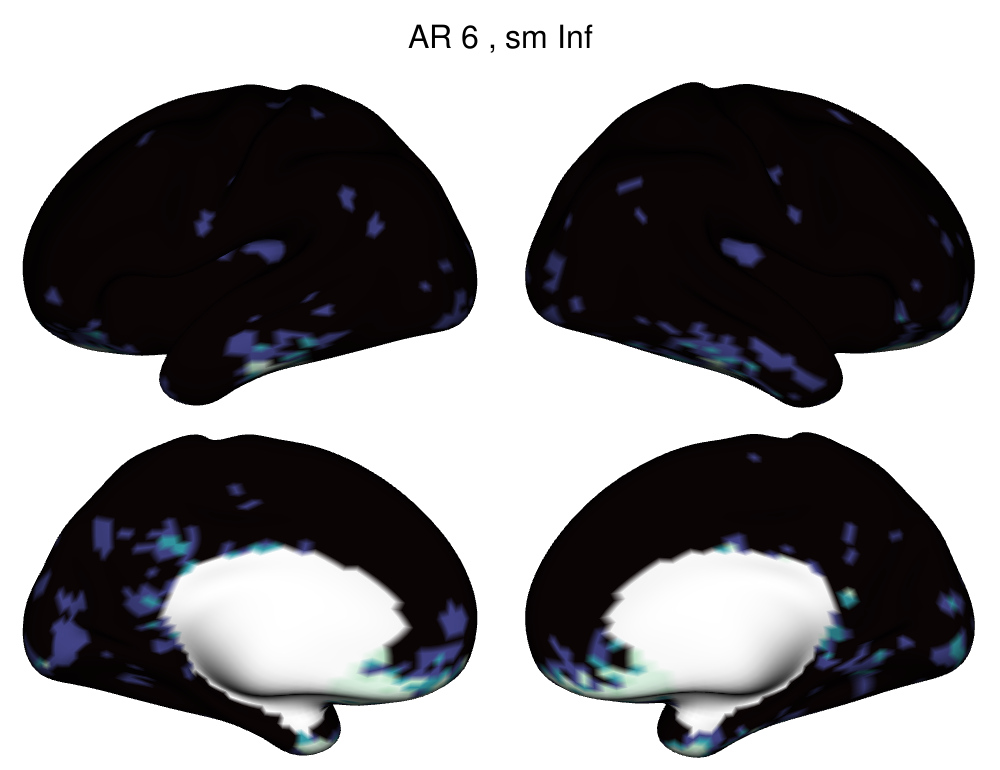}
    \end{tabular}
    \includegraphics[width=2in, trim = 0 7cm 0 0, clip]{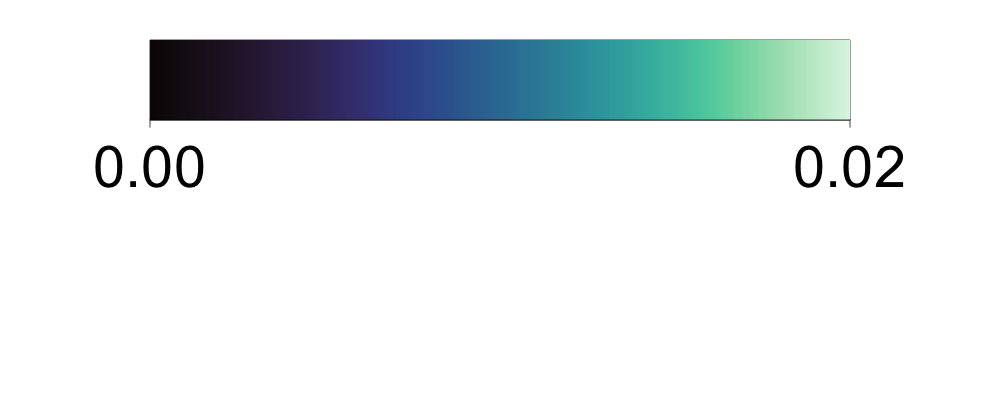}
    \caption{Proportion of sessions showing false positives at each vertex\\[10pt]}
\end{subfigure}
\begin{subfigure}{0.49\textwidth}
    \includegraphics[width=1\textwidth]{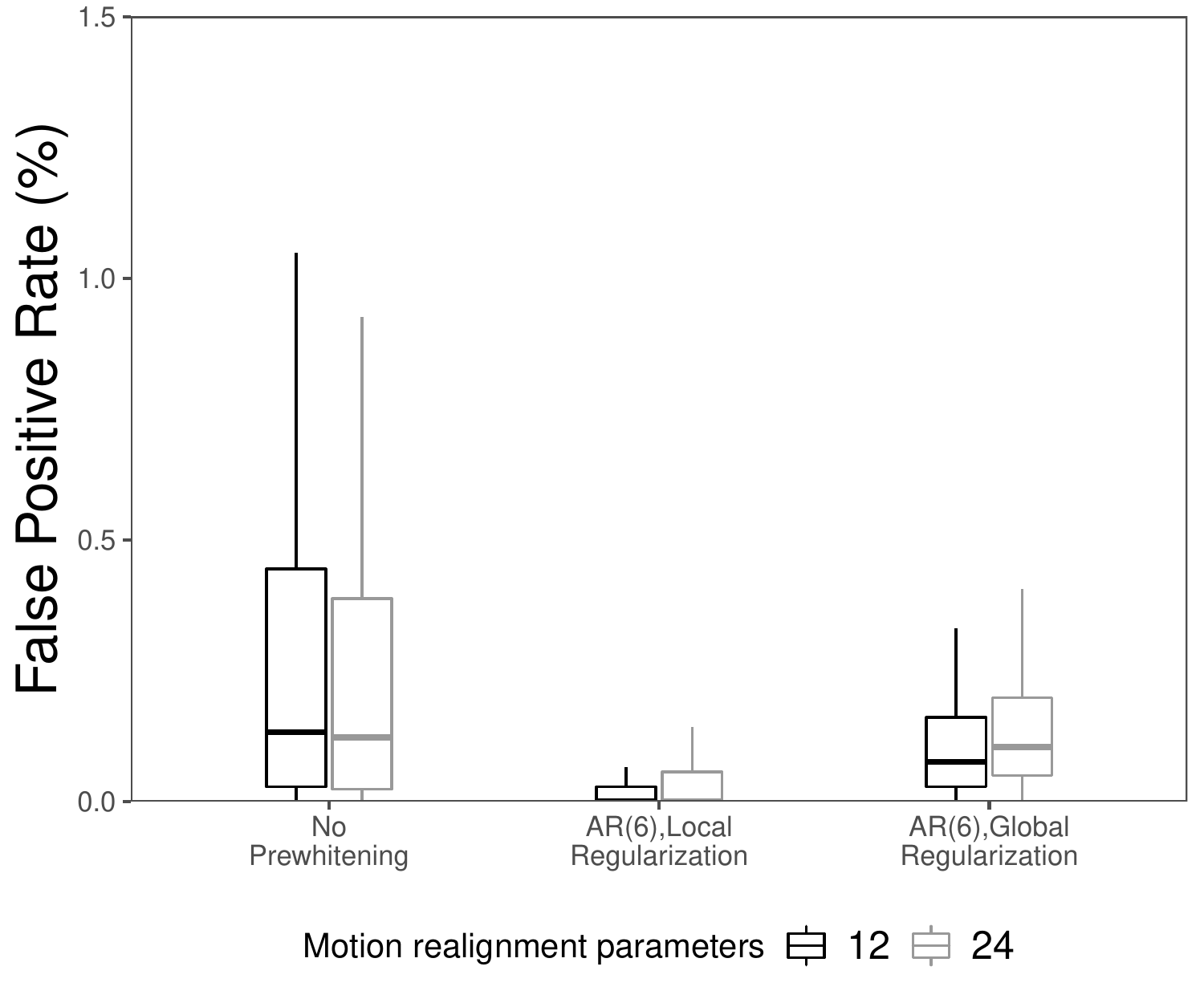}
    \caption{False positive rate (FPR)}
\end{subfigure}
\begin{subfigure}{0.49\textwidth}
    \includegraphics[width=1\textwidth]{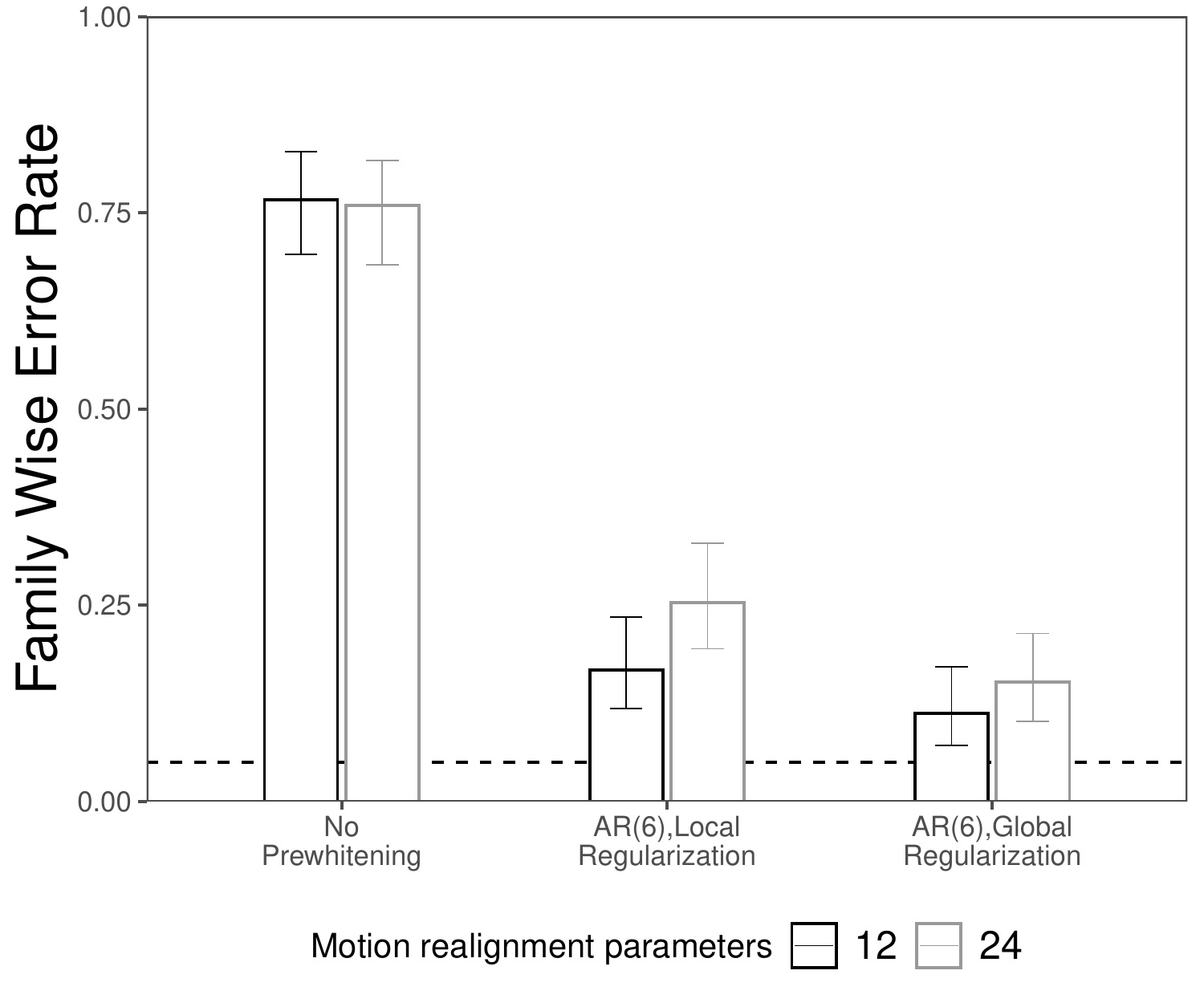}
    \caption{Family-wise error rate (FWER)} 
\end{subfigure}
\caption{\small \textbf{False positives in resting-state data before and after prewhitening.} (A) Values at each vertex represent the proportion of sessions where a false positive is detected. (B) Boxplots representing the distribution across all sessions and subjects of the FPR, defined as the proportion of vertices flagged as false positives for a given scan. (C) FWER with $95\%$ Agresti-Coull confidence intervals for proportions. Prewhitening dramatically reduces false positive rates and brings FWER close to the nominal rate of 0.05 (dashed line). Local regularization of prewhitening parameters achieves near-zero FPR across nearly all sessions. Interestingly, including additional motion covariates (24 versus 12) seems to worsen both FPR and FWER when used alongside prewhitening.}
\label{fig:FPR_FWER}
\end{figure}

In \textbf{Figure \ref{fig:FPR_FWER}}, we examine the effect of prewhitening on false positives in null (resting-state) data.  Assuming a false on-off 10s boxcar design, we fit a GLM and perform a $t$-test on the task coefficient at every vertex. We perform Bonferroni correction of the $p$-values to control the FWER at $0.05$. Note that while Bonferroni correction is often considered overly conservative for volumetric fMRI analyses involving hundreds of thousands of tests, here we have resampled the data to $6,000$ vertices per hemisphere, so the number of tests being performed is an order of magnitude less. We have previously observed that Bonferroni correction is not more conservative than permutation testing in this data. Panel (A) displays the proportion of sessions showing a false detection at each vertex when no prewhitening is performed or when an AR(6) model is used for prewhitening with local or global coefficient regularization. Panel (B) displays the false positive rate (FPR), the proportion of vertices labeled as active in each session, averaged across all subjects, sessions and tasks. Panel (C) displays the FWER, the proportion of sessions exhibiting a single false positive vertex. In (B) and (C) we also consider the effect of including additional motion parameters (24) versus the 12 included by default in our analyses. We observe that for both local and global regularization, the inclusion of additional motion parameters actually worsense the FPR and FWER. This is in line with the slight increase in ACI we observe when these parameters are included in combination with prewhitening (see \textbf{Figure \ref{fig:RP24_vs_RP12}}). Taken together, these results suggest that the loss in degrees of freedom associated with including superfluous covariates in the GLM worsen the performance of prewhitening.  This illustrates that overparameterized GLMs may actually result in inflated false positive rates, in addition to their well-known effect of reducing power to detect true effects. Comparing the FPR and FWER before and after prewhitening, we see that prewhitening drastically reduces the FPR within each session, and achieves FWER fairly close to the nominal rate of $0.05$. 

Interestingly, using global regularization achieves slightly lower FWER while achieving slightly worse but still very low FPR. This surprisingly strong performance of global regularization stands in contrast to its poor performance in our task fMRI-based analyses, displayed in Figures \ref{fig:prewhitening_ACI} and \ref{fig:prewhitening_LJB}. This may indicate some limitations of using resting-state fMRI as ``null'' data for evaluating false positive control in task fMRI. There are many features of task fMRI data that may not be reflected in resting-state fMRI data.  For example, mismodeling of the task-induced HRF can induce residual autocorrelation, as shown in \textbf{Figure \ref{fig:ACI_avg}}. Inclusion of HRF derivatives only partly accounts for the task-related differences in autocorrelation, as shown in \textbf{Figure \ref{fig:lmer_FE_task}}.

\section{Discussion}
\label{sec:discussion}

In this paper, we have made three primary advances in prewhitening in fMRI. First, we performed a comprehensive analysis to examine the spatial topology of autocorrelation across the cortex and identify the different factors driving autocorrelation. Second, we evaluated the efficacy of a range of AR-based prewhitening methods at eliminating autocorrelation and controlling false positives. We found that ``local'' prewhitening methods that account for spatial variability strongly outperform ``global'' methods where the same filter is applied to each voxel or vertex of the brain. Third, we developed a fast implementation of local prewhitening, available through the open-source \texttt{BayesfMRI} R package, that overcomes the computational challenges associated with performing prewhitening at thousands of locations.

\subsection{Variable autocorrelation across the cortex results in spatially differential false positive control}

Using a mixed effects modeling approach and test-retest data from the Human Connectome Project, we showed that autocorrelation varies markedly across the cortex and is influenced by task-related differences, modeling choices, acquisition factors, and population variability. As a result, the spatial topology of autocorrelation in each fMRI scan is unique. Given the spatial variability in autocorrelation, global prewhitening will result in differential levels of false positive control across the brain or cortex. And because the spatial topology of autocorrelation is unique to each fMRI scan, the topology of false positive control as well as power will likewise vary across fMRI scans, even within the same study. For example, one subject may be less likely to see a significant effect in a certain region compared with another subject in the same study, simply due to differences in autocorrelation in that region.  These results illustrate the importance of prewhitening techniques that capture the spatial variability in autocorrelation, in order to avoid differential false positive rates across the cortex or across the brain. 

Current prewhitening methods implemented in major fMRI software tools often use a global prewhitening approach. One likely reason for this is the computational efficiency of global prewhitening, since it requires a single $T\times T$ matrix inversion, unlike local prewhitening which requires $V$ such inversions. Likewise, the GLM coefficients can be estimated in a single matrix multiplication step with global prewhitening, whereas local prewhitening requires $V$ multiplications. Another seeming advantage of global regularization of the prewhitening parameters is the low sampling variability in the estimates of those parameters, though this comes as the cost of large biases for specific locations. Local prewhitening can lead to noisier estimates of the prewhitening parameters, though smoothing can help combat this. While previous work based on volumetric fMRI found smoothing to be detrimental because of mixing signals across tissue classes \citep{luo2020improved}, our use of surface-based smoothing largely avoids this limitation. Using cortical surface fMRI data also has the advantage of reduced dimensionality and the option to further reduce dimensionality through resampling without significant loss of spatial resolution. This lower dimensionality, combined with an implementation optimized for speed, makes our approach to local prewhitening quite feasible (approximately 1 minute per run for the task fMRI we analyze here).

\subsection{Implications for volumetric fMRI analyses}
\label{sec:discussion_vol}

Our analysis focused on cortical surface-based analysis, but our findings have major implications for volumetric fMRI analysis as well.  The issue of spatially varying autocorrelation is actually more salient in volumetric fMRI, because autocorrelation is known to differ markedly across tissue classes, with CSF generally exhibiting higher autocorrelation and white matter exhibiting lower autocorrelation \citep{bollmann2018serial, luo2020improved}. A global prewhitening approach in volumetric fMRI may have more severe consequences than in surface-based analyses because of the more dramatic differences in autocorrelation across tissue classes. For example, to eliminate autocorrelation within CSF and thereby control false positive rates there, we may over-whiten within gray matter. Even if we target the gray matter, standard volumetric smoothing exacerbates differences in autocorrelation, increasing autocorrelation in voxels near CSF and decreasing autocorrelation in voxels near white matter. While these issues point to the importance of local prewhitening in volumetric fMRI analysis, the higher dimensionality of that data introduces new computational challenges.  For example, our implementation of local prewhitening would take approximately $10$ minutes per run for a volumetric analysis involving 100,000 voxels or more, compared with $1$ minute per run for 12,000 surface vertices. Additionally, our local regularization approach of smoothing AR coefficients may not translate seamlessly to volumetric fMRI analysis, given the risks associated with smoothing across tissue classes. For volumetric fMRI analysis, it may be preferable to avoid AR coefficient smoothing or employ smoothing techniques that respect tissue class boundaries.

\subsection{Low-order AR models perform surprisingly well when allowed to vary spatially}

We were somewhat surprised by the fairly strong performance of AR(1) models when the AR coefficients were allowed to vary spatially through local regularization. This echoes the relatively strong performance of AFNI \citep{cox1996afni}, which assumes a spatially varying ARMA(1,1) model with no smoothing, observed by \cite{olszowy2019accurate}. In our analysis, locally regularized AR(1) prewhitening consistently outperformed globally regularized AR(6) prewhitening at reducing autocorrelation.  We generally saw the best performance for local AR(6) prewhitening, but its improvement over local AR(1) was small compared to the difference between local and global regularization. This suggests that for volumetric fMRI analysis where the computational burden associated with voxel-specific prewhitening may be substantial, lower-order AR models may be worth consideration since they would speed up estimation of the prewhitening coefficients and matrix inversion. 

\subsection{FWER is not the whole picture}

Because many multiplicity correction methods focus on controlling the family-wise error rate (FWER) or the probability of observing a single false positive voxel, vertex or cluster, FWER control is often used as an evaluation metric for prewhitening methods. While this can be a useful metric in determining whether important modeling assumptions (e.g., independent residuals) have been satisfied, it is not the only important one. We advocate for two additional considerations: equal false positive control across the brain or cortex, and avoiding unnecessary loss of power.  Global prewhitening that does not consider the spatial variance in autocorrelation may control the FWER, but will generally fail to achieve spatially homogeneous levels of false positive control. As a result, we may be much more likely to observe false positives in certain regions, which may also differ across subjects. In addition, global prewhitening will tend to over-whiten regions with low autocorrelation, which can lead to overly conservative inference, manifesting as a lack of power to detect effects. Just as we will be more likely to see false positives in certain regions, other regions will suffer disproportionately from a loss of power to detect real effects. Indeed, achieving nominal FWER without spatially homogeneous false positive control will almost surely come at the cost of unnecessary loss of power in many parts of the brain.

% \subsection{Resting-state fMRI may not accurately reflect false positive rates in task fMRI}

% Limitations of using resting-state fMRI to assess prewhitening efficacy

\subsection{Acquisition-induced distortions change the spatial topology of autocorrelation and false positive control}

One striking finding in our analysis was the effect of phase encoding direction on the spatial topology of residual autocorrelation. Comparing the RL and LR phase encoding directions, LR generally produced higher autocorrelation in the right lateral cortex and left medial cortex, while RL generally produced higher autocorrelation in the left lateral cortex and the right medial cortex. In other words, the choice of phase encoding direction generally had opposing effects on the lateral and medial cortices within each hemisphere, as well as across hemispheres. Why might this be? The LR and RL phase encoding directions are known to introduce lateralized distortions, even after distortion correction. These distortions are due to areas of varying magnetic susceptibility giving rise to signal stretching and signal pile-up based on the direction of the phase-encode direction for acquisitions such as echo-planar imaging \citep{jezzard1999sources}. Those distortions cause a slight misalignment of the fMRI data on the structure of the brain. This has the result of mixing CSF signals with higher autocorrelation into some cortical areas, and mixing white matter signals with lower autocorrelation into others. For example, the higher autocorrelation on the right lateral cortex with the LR acquisition may result from the introduction of CSF signals produced by a slight shift of gray matter voxels to the left; a similar shift to the left of the right medial cortex will introduce white matter signals, resulting in lower autocorrelation. This is somewhat analogous to the effect of volumetric smoothing on gray matter voxels bordering CSF and white matter observed by \citep{luo2020improved} (though note that these effects of volumetric smoothing would be in addition to, and perhaps exacerbate, the effects of distortions). As a result, in the HCP there may be sizeable discrepancies in false positive control and power across and within each hemispheres before prewhitening or when using global prewhitening.  While such lateralized effects are somewhat unique to HCP-style acquisitions that employ left-to-right or right-to-left phase encoding, other acquisitions can introduce different types of distortions that may also change the spatial topology of autocorrelation. Different acquisitions may therefore produce very different spatial distributions of false positive control, if not addressed through an effective local prewhitening strategy.

\subsection{Limitations and future directions}

Our study has several limitations. First, our analysis was based on a single dataset, the Human Connectome Project, which includes data from healthy young adults collected using a certain multi-band acquisition protocol.  Previous studies have found the baseline level of autocorrelation, as well as the efficacy of prewhitening, to differ across datasets of varying TR \citep{corbin2018accurate, olszowy2019accurate}. Future work should examine the generalizability of our findings, particularly the efficacy of our local prewhitening approach, to data collected with different acquisition protocols and in more diverse populations. Similarly, a valuable area of future work would be to assess the efficacy of local prewhitening strategies in volumetric task fMRI data, which presents unique challenges as described in Section \ref{sec:discussion_vol}. 

Our study, as well as most prior studies on the efficacy of prewhitening in task fMRI analyses, focused on the ability of prewhitening techniques to effectively mitigate autocorrelation and control false positives. Here we discussed, but did not explicitly analyze, the possibility of a loss of power due to over-whitening.  We argue that this is more likely with global prewhitening strategies that aim to achieve false positive control within a particular region or tissue class, at the risk of over-whitening in other areas. A valuable area of future work would be to examine the effect of different prewhitening techniques on power across the brain.

Here, we considered the effect of HRF modeling strategy on residual autocorrelation, and observed that the inclusion of temporal and dispersion derivatives of the HRF helped to alleviate it. However, we did not consider alternative, potentially advantageous HRF modeling strategies such as the inverse logit or finite impulse response models \citep{lindquist2009modeling}. When longer scan durations are available, these more flexible models may help account for additional spatial, within-subject and between-subject heterogeneity in HRF shape, onset and duration, particularly in more diverse populations. Thus, these more flexible HRF modeling strategies could further reduce autocorrelation and its spatial variance.

Finally, one limitation of our implementation of AR-based prewhitening implementation is that we did not account for potential bias in the prewhitening matrix due to using the fitted residuals as a proxy for the true residuals. Since the fitted residuals have a different dependence structure induced by the GLM, their covariance matrix is not equal to that of the true residuals. This bias will generally be worse in overparameterized GLMs, which may help explain why we observed a slightly detrimental effect of including all 24 motion regressors when prewhitening was also performed (see Supplementary Figure \ref{fig:RP24_vs_RP12}). A valuable topic of future work would be to develop prewhitening methods that formally model and adjust for this source of bias.

\section{Conclusion} 

We performed a comprehensive investigation of the sources and patterns of residual autocorrelation across the cortical surface in multi-band task fMRI data.  Our analysis revealed dramatic spatial differences in autocorrelation across the cortex. This spatial topology is unique to each session, being influenced by the task being performed, the acquisition technique, various modeling choices, and individual differences. If not accounted for, these differences will result in differential false positive control and power across the cortex and across subjects. We evaluated the efficacy of different prewhitening methods to mitigate autocorrelation and control false positives.  Our findings demonstrate that allowing the prewhitening filter to vary spatially is crucial to effectively reducing autocorrelation and its spatial variability across the cortex. Our computationally efficient implementation of ``local'' prewhitening is available in the open-source \texttt{BayesfMRI} R package.

\section*{Funding}
This work was supported by the National Institute of Biomedical Imaging and Bioengineering at the National Institutes of Health (NIH) under grant R01EB027119, by the National Institute of Neurobiological Disorders and Stroke at the NIH under the grant R25NS117281, and by the National Science Foundation (NSF) under grant IIS-2023985 from the Division of Information and Intelligent Systems.

\bibliography{mybib.bib}
\bibliographystyle{apalike}

\appendix

\renewcommand\thefigure{\thesection.\arabic{figure}}
\setcounter{figure}{0}

\newpage
\setcounter{page}{1}

\section{Supplementary Figures}

\begin{figure}[H]
    \centering
    \begin{tabular}{ccc}
    Average ACI,  & Average ACI, & Difference  \\
    12 RPs & 24 RPs  & (24 RPs - 12 RPs) \\
    \includegraphics[width=1.5in, trim=0 0 7in 0, clip]{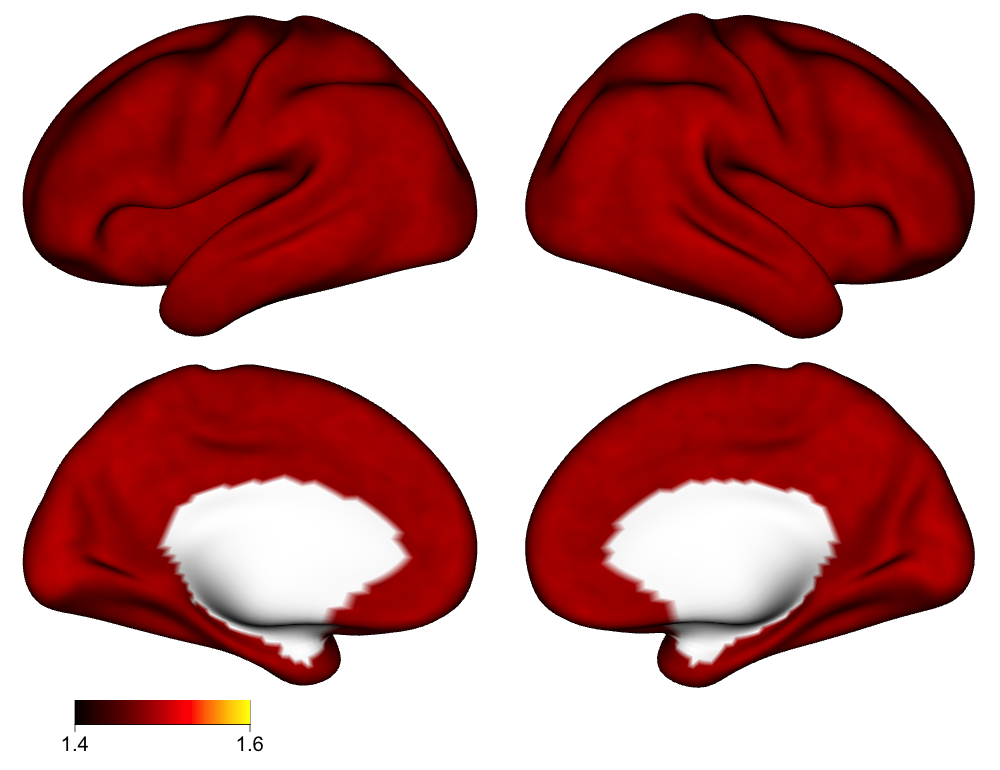} &
    \includegraphics[width=1.5in, trim=0 0 7in 0, clip]{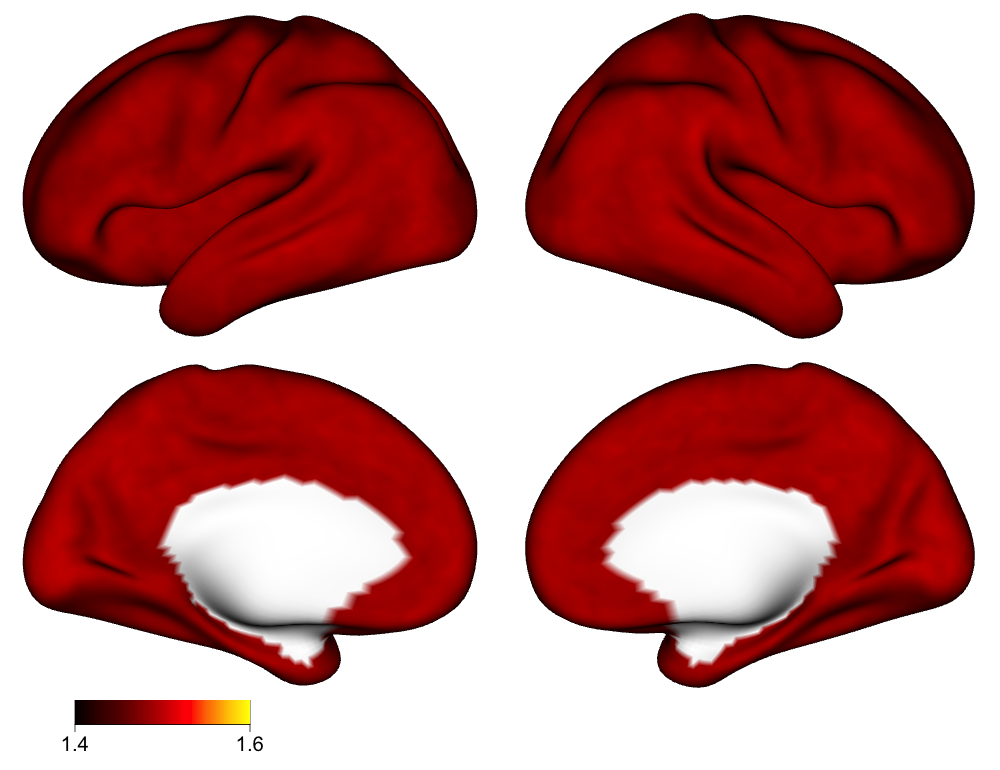} &
    \includegraphics[width=1.5in, trim=0 0 7in 0, clip]{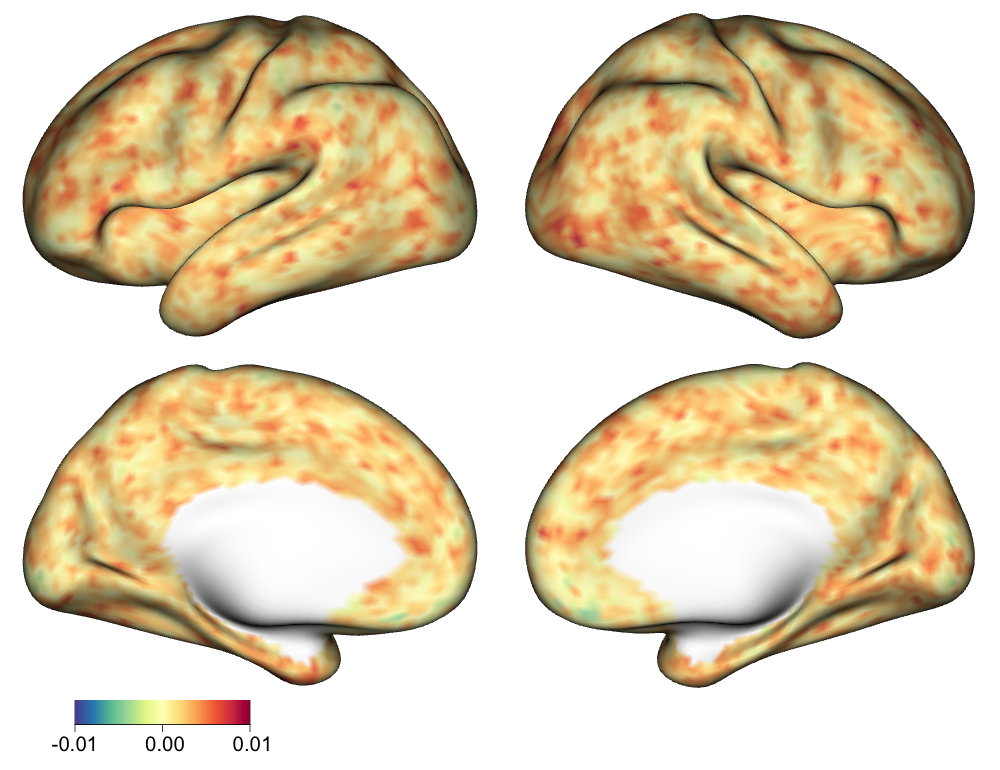} 
    \end{tabular}
    \caption{\small\textbf{Effect of including additional motion covariates on autocorrelation, when effective prewhitening is performed.} For prewhitening, we use an AR(6) model with local regularization of AR model coefficients, which we observe to be highly effective at reducing autocorrelation. The first two columns show the average autocorrelation index (ACI) across all subjects, sessions and tasks when 12 realignment parameters (RPs) or 24 RPs are included in each GLM.  The last column shows the difference (24 RPs - 12 RPs).  The mostly warm colors indicate that using additional RPs results in very slightly \textit{worse} autocorrelation when effective prewhitening is performed.}
    \label{fig:RP24_vs_RP12}
\end{figure}

\section{Prewhitening Algorithm}\label{app:PWalgo}

%TO appendix:
%We begin finding the prewhitening matrix $\bfL_v$ by fitting the GLM in equation (\ref{eqn:GLM}) to each vertex, and then estimating the $p$ AR coefficients for the residuals of the model using the Levinson-Durbin recursion \citep{Brockwell_Davis}. Based on the estimated AR coefficients $\phi_{v,1},\dots,\phi_{v,p}$ and white noise variance $s_v$ after global or local regularization, the temporal precision matrix $\bfW_v$ is created as a sparse band matrix. The prewhitening matrix $\bfL_v$, satisfying the relation $\bfW_v = \bfL_v\bfL_v'$, is found, where $\bfL_v = \mathbf{U}_v\mathbf{D}_v\mathbf{U}_v'$, $\mathbf{U}_v$ is the matrix of eigenvectors of $\bfW_v$, and $\bfD_v$ is the diagonal matrix of the square roots of the eigenvalues of $\bfW_v$. 

\begin{algorithm}[H]
	\SetAlgoLined
	\KwResult{Prewhitened data values for the response and design across all vertices $v = 1,\ldots,V$}
	\For{$v \gets 1$ \KwTo $V$}{
	    Find estimates $\hat{\bfbeta}_v$ by fitting the GLM at vertex $v$\;
	    Find AR coefficients $\phi_{v,1},\dots,\phi_{v,p}$ and white noise variance $s_v$ after global or local regularization for the time series $\bfy_v - \bfX\hat{\bfbeta}_v$ using the Levinson-Durbin recursion \citep{Brockwell_Davis}\;
	    \textit{Create the symmetric temporal precision matrix $\bfQ_v^{-1}$ as a sparse band matrix taking the value 1 on the diagonal and $-\phi_{v,q}$ for the $q$th location off the diagonal\;
	    Set $\bfS_v^{-1} = ((1 / s_v) \bfI_T) \bfQ_v^{-1}$\;
	    Find the eigenvectors and eigenvalues of $\bfS_v^{-1}$, where  $\mathbf{U}_v$ is the matrix of eigenvectors and $\mathbf{D}_v$ is the diagonal matrix of corresponding eigenvalues\;
	    Compute the prewhitening matrix $\bfW_v = \mathbf{U}_v\mathbf{D}_v\mathbf{U}_v'$\;
	    Set off-diagonal values $W_{ij} =0 $ at row $i$ and column $j$ within $\bfW_v$ if $|i - j| > p$\;}
	}
	Create $\bfW$ as a block-diagonal matrix for all prewhitening matrices $\bfW_v$\;
	Output prewhitened response $\tilde{\bfY} = \bfW \bfY$ and block-diagonal prewhitened design matrix $\tilde{\bfX}$ with elements $\bfW_v\bfX$\;
	\caption{Prewhitening time series data in order to reduce temporal dependence in residuals. Steps in \textit{italics} are performed using C++ for computational efficiency.}
\end{algorithm}

\end{document}